\documentclass[useAMS,usenatbib]{mn2e}
\usepackage{graphicx,amsmath,amssymb,ifthen}

\newboolean{astroph}
\setboolean{astroph}{true}

\newboolean{mn2e}
\setboolean{mn2e}{false}

\newboolean{aastex}
\setboolean{aastex}{false}

\newcommand{\Msun}{\ensuremath{\mathrm{M}_\odot}}

\newcommand{\LCDM}{\ensuremath{\Lambda\mathrm{CDM}}}
\newcommand{\Msunpyr}{\ensuremath{\mathrm{M}_\odot\mathrm{yr}^{-1}}}

\newcommand*{\dif}{\ensuremath{\mathrm{d}}}

\newcommand{\kmps}{\ensuremath{\mathrm{km\ s}^{-1}}\ }

\newcommand{\Mpc}{\ensuremath{\mathrm{Mpc}}}

\newcommand{\Mpch}{\ensuremath{h^{-1}}\Mpc}

\newcommand{\h}{\ensuremath{h}}
\newcommand{\mun}{\ensuremath{\langle\mu_n\rangle}}
\newcommand{\mum}{\ensuremath{\langle\mu_m\rangle}}

\newcommand{\Reff}{\ensuremath{R_{\mathrm{eff}}}}

\newcommand{\figMfunc}{
  \begin{figure}
    \includegraphics[width=3in]{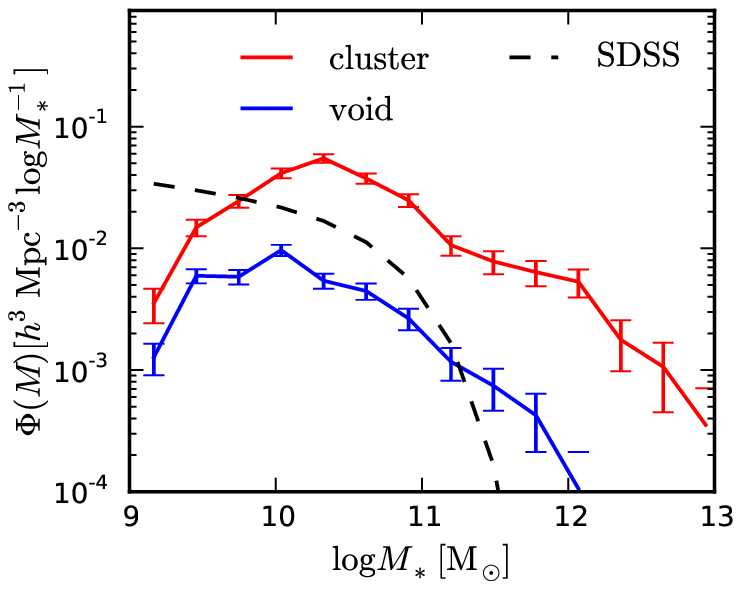}
    \caption{{\rm The $z=0$ mass function of the cluster and void box shown
      in red and blue, respectively. The mass function from SDSS is
      shown by the black dashed line \citep{Li2009}. The mass
      functions of the simulated boxed bracket the observations for
      $10^{10}\ \Msun < M_* < 10^{11} \Msun$. The simulations overproduce
      galaxies more massive than $10^{11}\ \Msun$. }}
    \label{fig:massFunc}
  \end{figure}
}
\newcommand{\figaccrfrac}{
  \begin{figure}
    \includegraphics[width=3in]{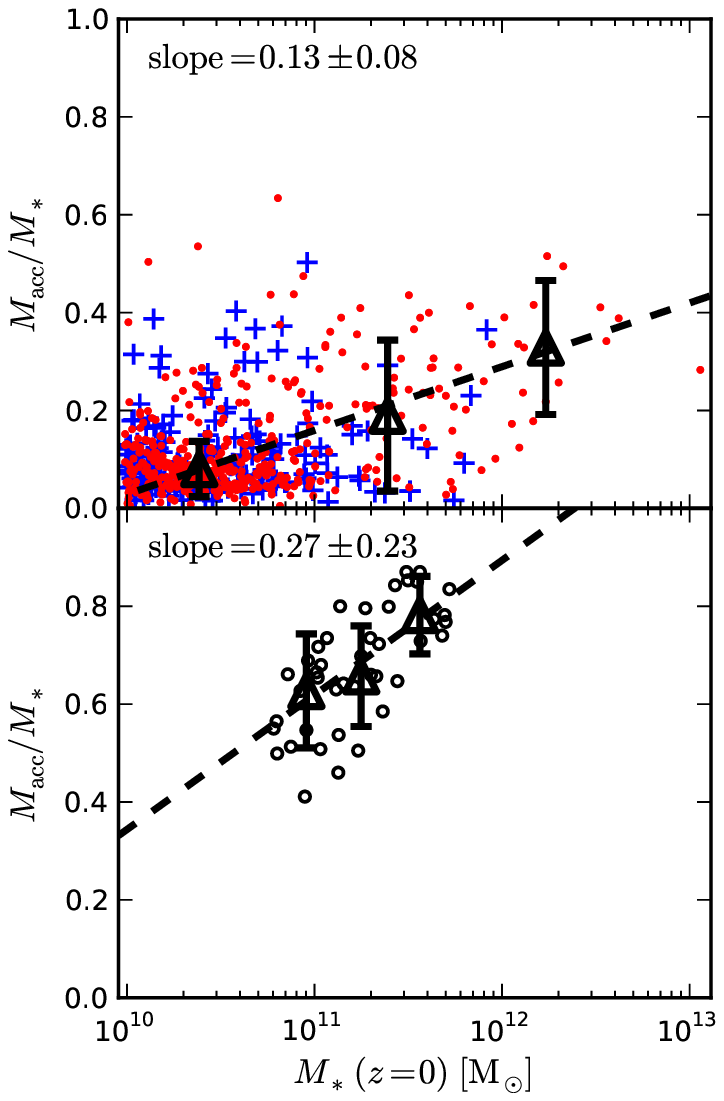}
    \caption{The fraction of accreted material ($M_{acc}/M_*$) as a
      function of final ($z=0$) stellar mass. In the top panel, the blue(red)
      points are for the void(cluster) box. The large black triangles are
      the medians and inter-quartile ranges in $3$ mass bins. The line
      is fit 
      through 
      the medians. The lower panel shows the same for the
      galaxies simulated in Os10 (data from L. Oser). {\rm The
        accreted fraction found in this work is significantly smaller
        than that in Os10 for very massive galaxies.}}
    \label{fig:m_macc}
  \end{figure}
}
\newcommand{\figaccisages}{
  \begin{figure}
    \includegraphics[width=3in]{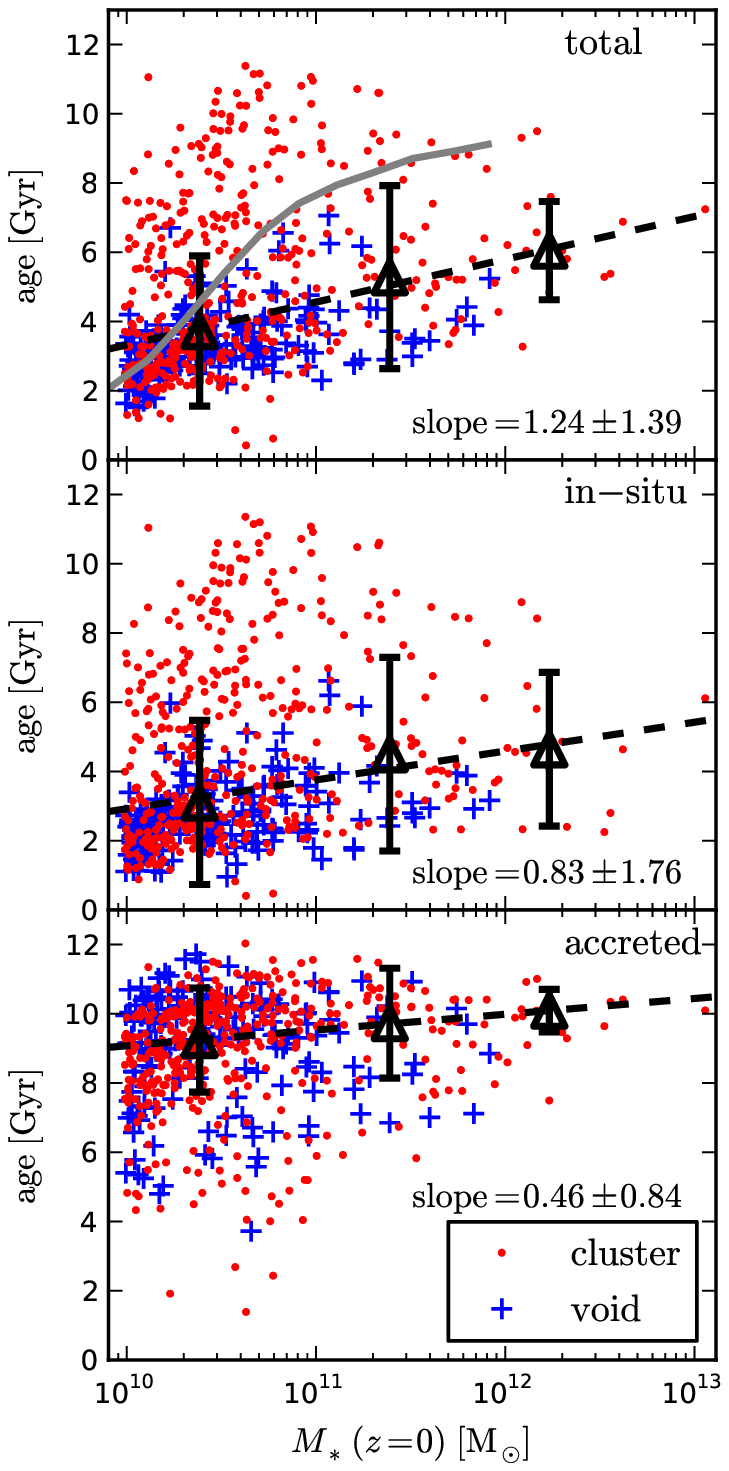}
    \caption{{\rm The mean (luminosity-weighted) stellar ages for all
        the stars (top), and the in-situ (middle) and accreted (bottom)
        stars as a function of total galaxy stellar mass.  The large
        black triangles are 
      the medians and inter-quartile ranges in $3$ mass bins. The
      dashed line is fit the medians. The gray solid line in the top
      panel is the median observed mass--age relation from SDSS
      \citep{Gallazzi2005}. The galaxies in the simulation
      are young compared to observed galaxies. Nonetheless, the
      accreted stars are uniformly older than the in-situ stars.} }
    \label{fig:m_agesISACC}
  \end{figure}
}
\newcommand{\figaccismetals}{
  \begin{figure}
    \includegraphics[width=3in]{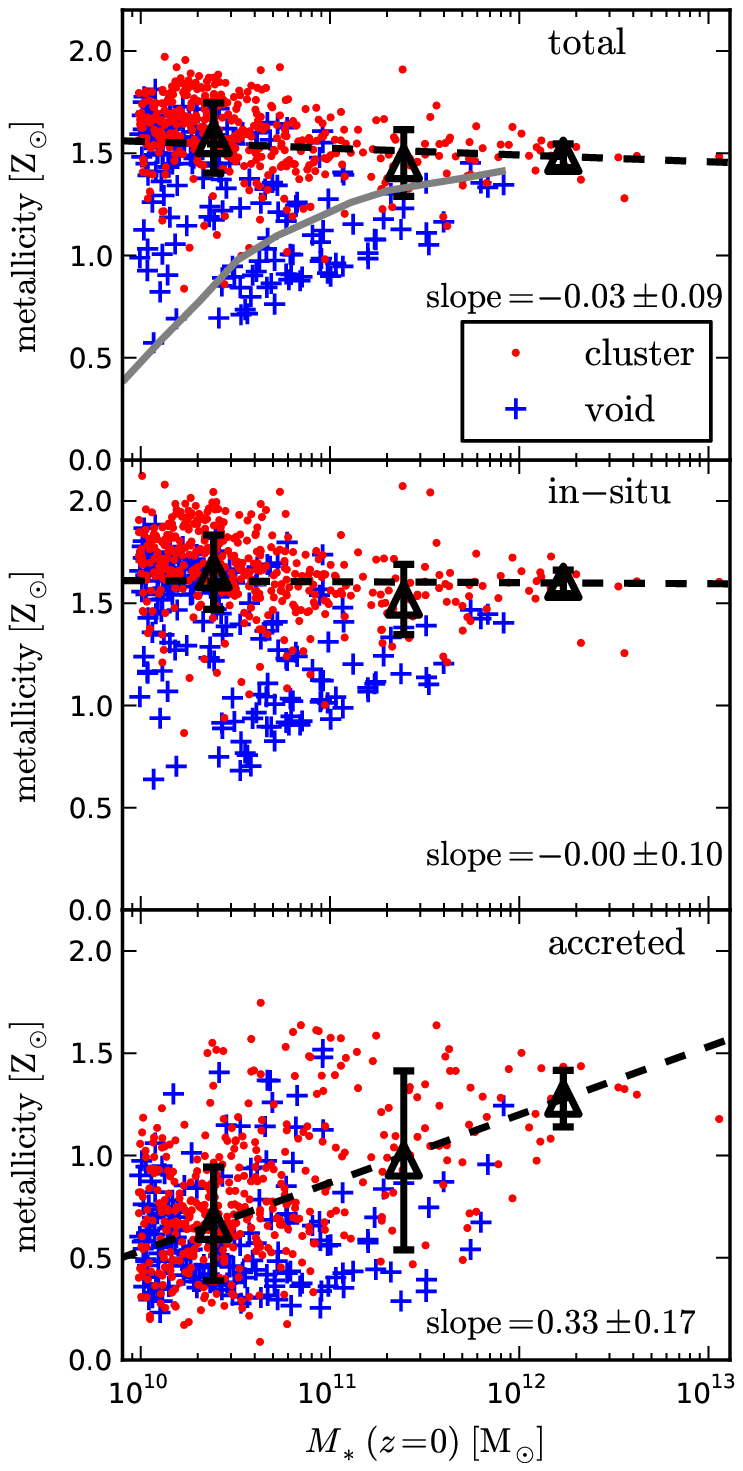}
    \caption{{\rm The mean (luminosity-weighted) stellar metallicities
        for all the stars (top), and the stars divided into in-situ
        (middle) and accreted (bottom) stars as a function of total
        galaxy stellar mass. The large black triangles are
      the medians and inter-quartile ranges in $3$ mass bins. The
      dashed line is fit the medians. The gray solid line in the top
      panel is the median observed mass--metallicity relation from SDSS
      \citep{Gallazzi2005}. The galaxies in the simulation
      show no mass metallicity relation, although the maximum
      metallicity is consistent with observations.  The
      symbols are as in Figure \ref{fig:m_agesISACC}.}} 
    \label{fig:m_zsolISACC}
  \end{figure}
}
\newcommand{\figAgeMassGrp}{
  \begin{figure}
    \includegraphics[width=3in]{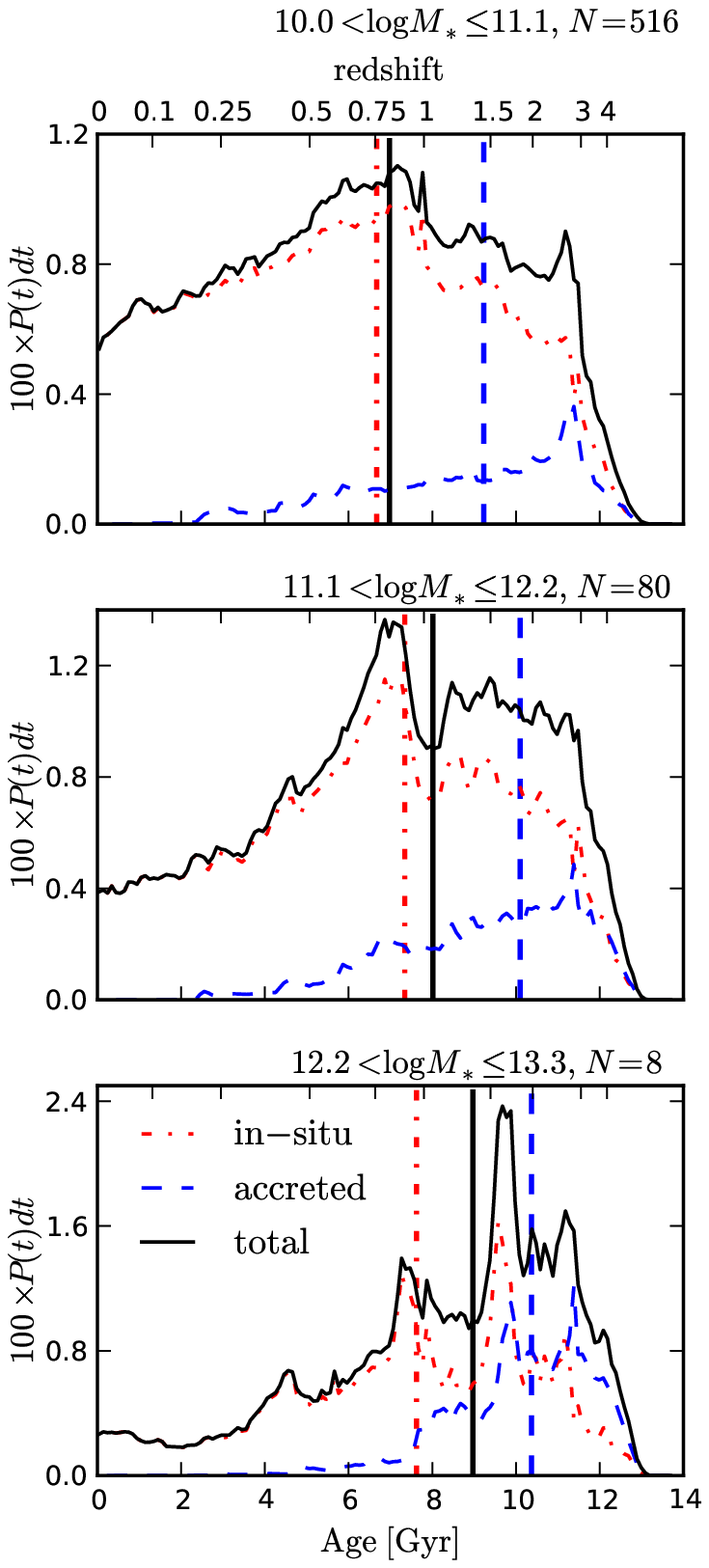}
    \caption{The probability distribution of stellar ages for galaxies
      grouped by mass. The black solid line is the total stellar age
      distribution. The blue dashed line is the accreted stars and the
      red dash-dotted line is the in-situ stars. The similar vertical
      lines show the median age of the stars in each category. {\rm For all
    mass bins, the accreted stars are $\sim 2$ Gyrs older than the
    in-situ stars.}}
    \label{fig:mGroup_ages}
  \end{figure}
}
\newcommand{\figZMassGrp}{
  \begin{figure}
    \includegraphics[width=3in]{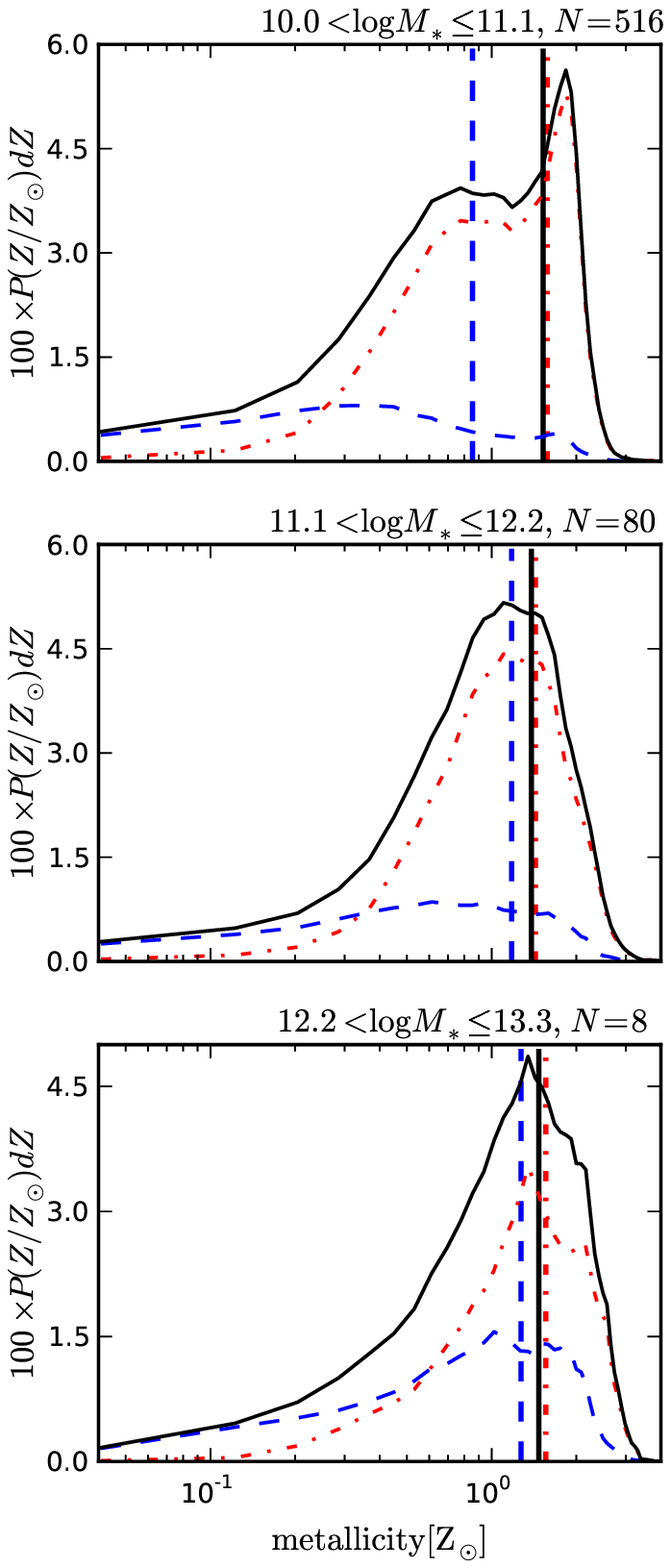}
    \caption{The probability distribution of stellar metallicities for
      galaxies grouped by mass. The lines are as in Figure \ref{fig:mGroup_ages}. {\rm At all masses, the accreted stars are more metal poor than in-situ stars. }}
    \label{fig:mGroup_metals}
  \end{figure}
}
\newcommand{\figBuildMassGrp}{
  \begin{figure}
    \includegraphics[width=3in]{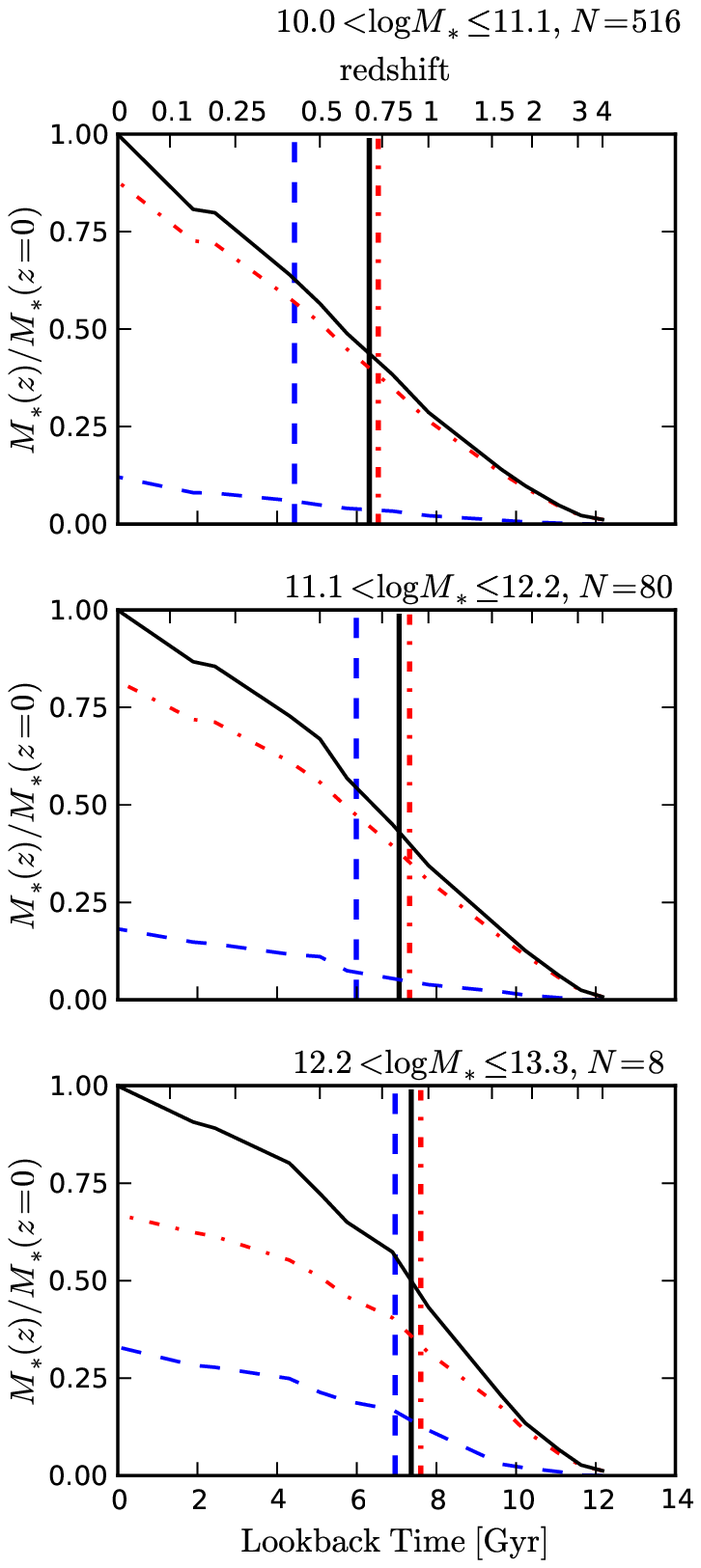}
    \caption{The stellar mass assembly histories for galaxies in three
      mass bins. The lines are the same as in
      Fig. \ref{fig:mGroup_ages}. The vertical lines show the time at
      which $50\%$ of the in-situ, accreted, and total mass is in place. {\rm While the accreted stars are typically formed at $z\sim 1.5$, they are assembled at $z \sim 0.6$.}}
    \label{fig:mGroup_assemble}
  \end{figure}
}
\newcommand{\figProfileMass}{
  \begin{figure}
    \includegraphics[width=3in]{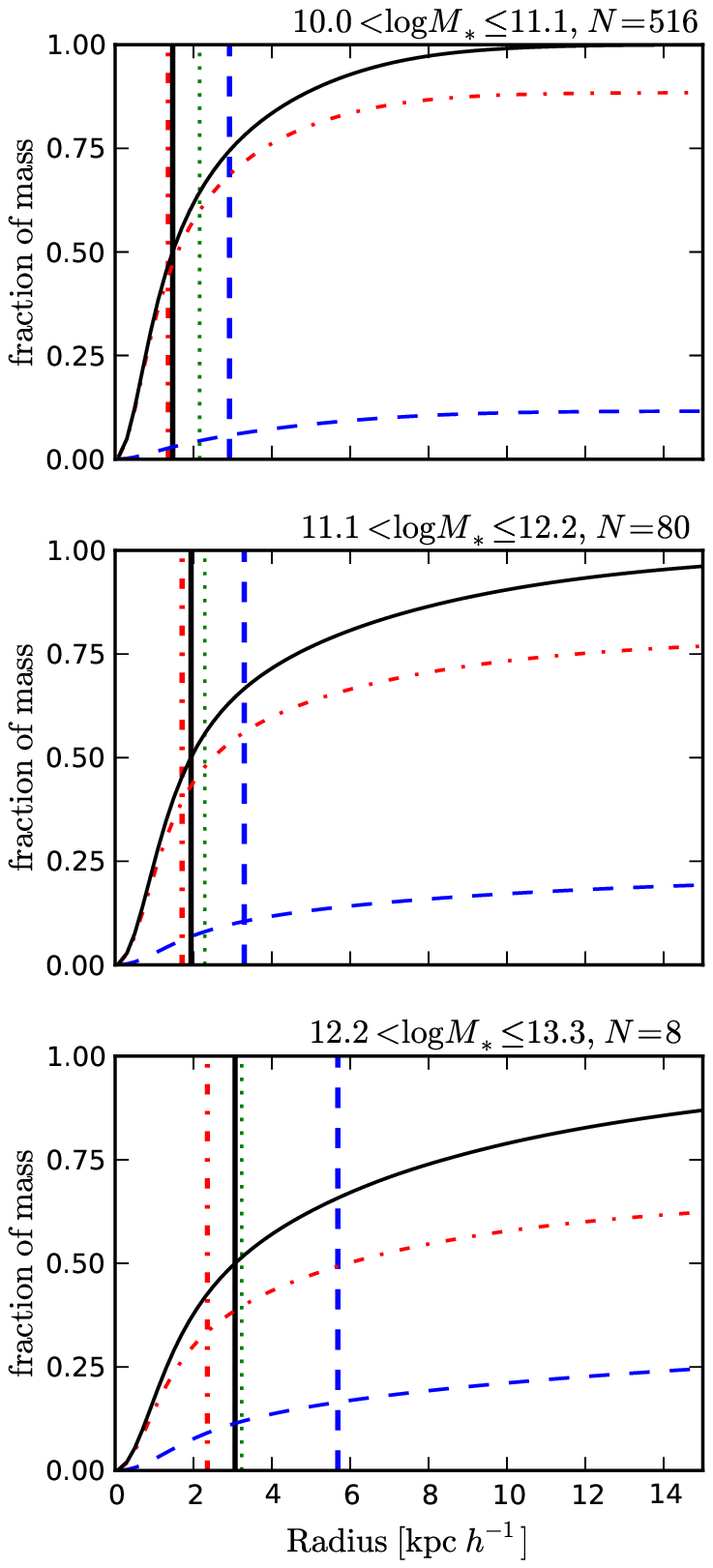}
    \caption{The cumulative, 3-dimensionally averaged radial stellar mass profiles for
      the galaxies in three mass bins. The lines are the same as in
      Figure \ref{fig:mGroup_ages}. The vertical lines show the
      3-dimensional half-mass 
      radii for the in-situ, accreted and
      total stellar mass profiles. {\rm The green dotted lines show the
      half-mass radius for material accreted from resolved galaxies
      only. These represent the minimum radius for the accreted
      stars. The accreted stars are consistently found at larger radii
    than the stars formed in-situ.} }
    \label{fig:mGroup_profile}
  \end{figure}
}
\newcommand{\figEnvAge}{
  \begin{figure}
    \includegraphics[width=3in]{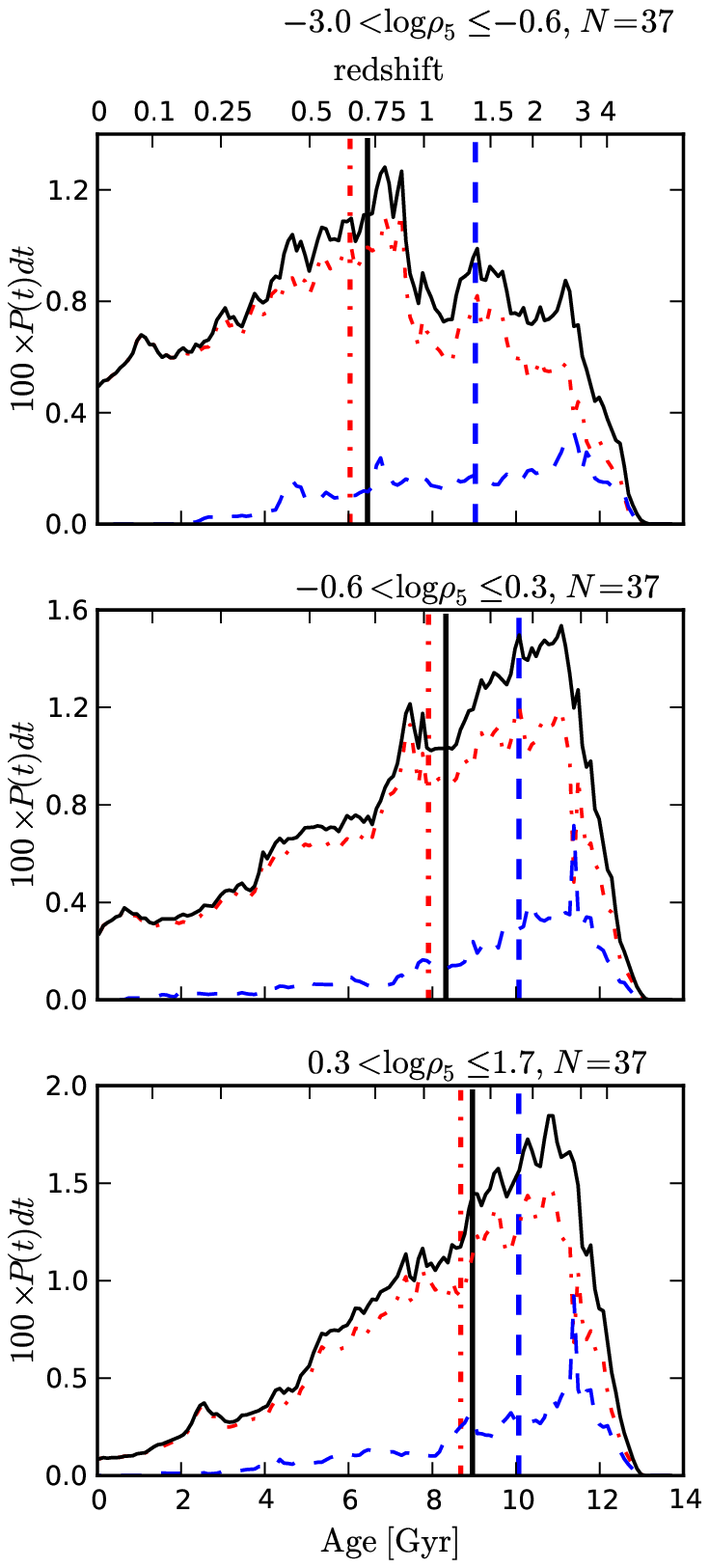}
    \caption{The probability distribution of stellar ages for galaxies
      grouped by environment ($\rho_5$). The galaxies all have stellar
      masses between $6\times 10^{10}\Msun < M_* < 3\times 10^{10}\Msun$. The
      lines are the same as in Figure \ref{fig:mGroup_ages}. {\rm Although
      the accreted fraction is only a weak function of environment,
      the median stellar age of the in-situ stars is a strong function of
    environment. }}
    \label{fig:envGroup_age}
  \end{figure}
}
\newcommand{\figEnvNormAge}{
  \begin{figure}
    \includegraphics[width=3in]{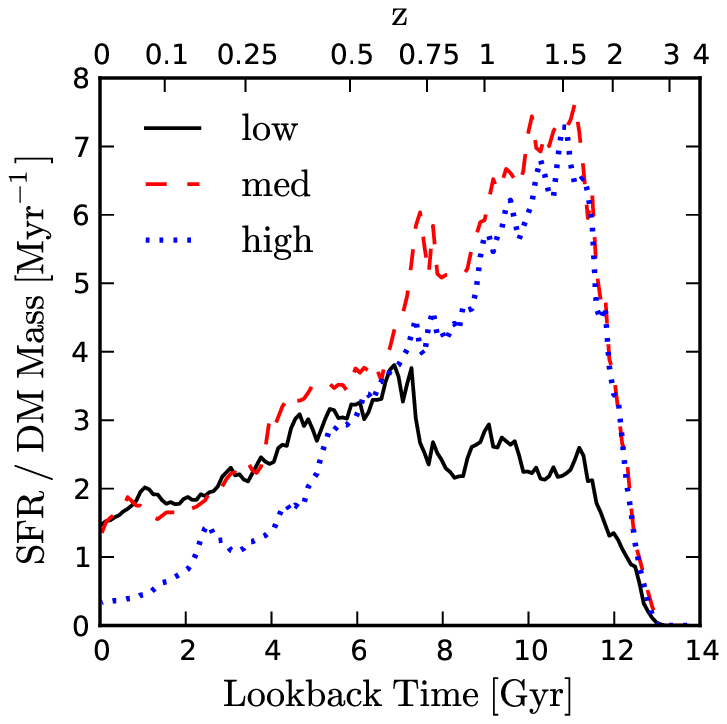}
    \caption{The star formation rates in different environments, for
      galaxies with stellar masses $6\times 10^{10}\Msun < M_* <
      3\times 10^{11}\Msun$ at $z=0$. The different curves are
      normalized to the average dark matter mass associated with each
      galaxy, as given by $1/5$ of the dark matter mass to the fifth
      nearest neighbor. {\rm At early times, the star formation occurs
      in the densest regions, while at late times the field galaxies
      dominate the star formation.}}
    \label{fig:envNormed_age}
  \end{figure}
}
\newcommand{\figmergerratio}{
  \begin{figure}
    \includegraphics[width=3in]{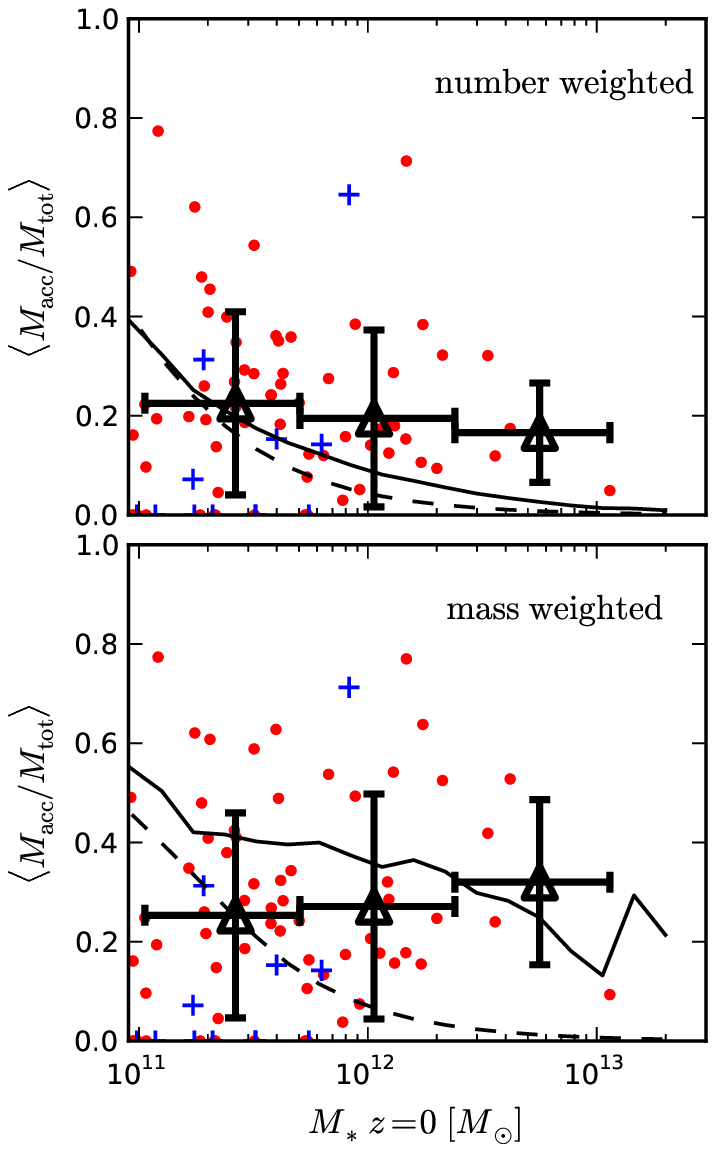}
    \caption{The mean merger ratios as a function of stellar mass. The
      red points are from the cluster box and the blue crosses are
      from the void box. The top panel shows the number-weighted mean
      merger ratio,
      while the bottom panel shows the mass-weighted mean merger
      ratio. The solid line is the expected mean merger ratio computed using
      the {\rm unweighted} mass function from the simulation. The
      dotted line shows the expected mean  
      merger ratio for a Schechter function galaxy mass
      distribution with$\alpha=-1.16$ and $\log M_* = 10.8$.}
    \label{fig:m_mergerratio}
  \end{figure}
}
\newcommand{\figmhostz}{
  \begin{figure*}
    \includegraphics[width=6.4in]{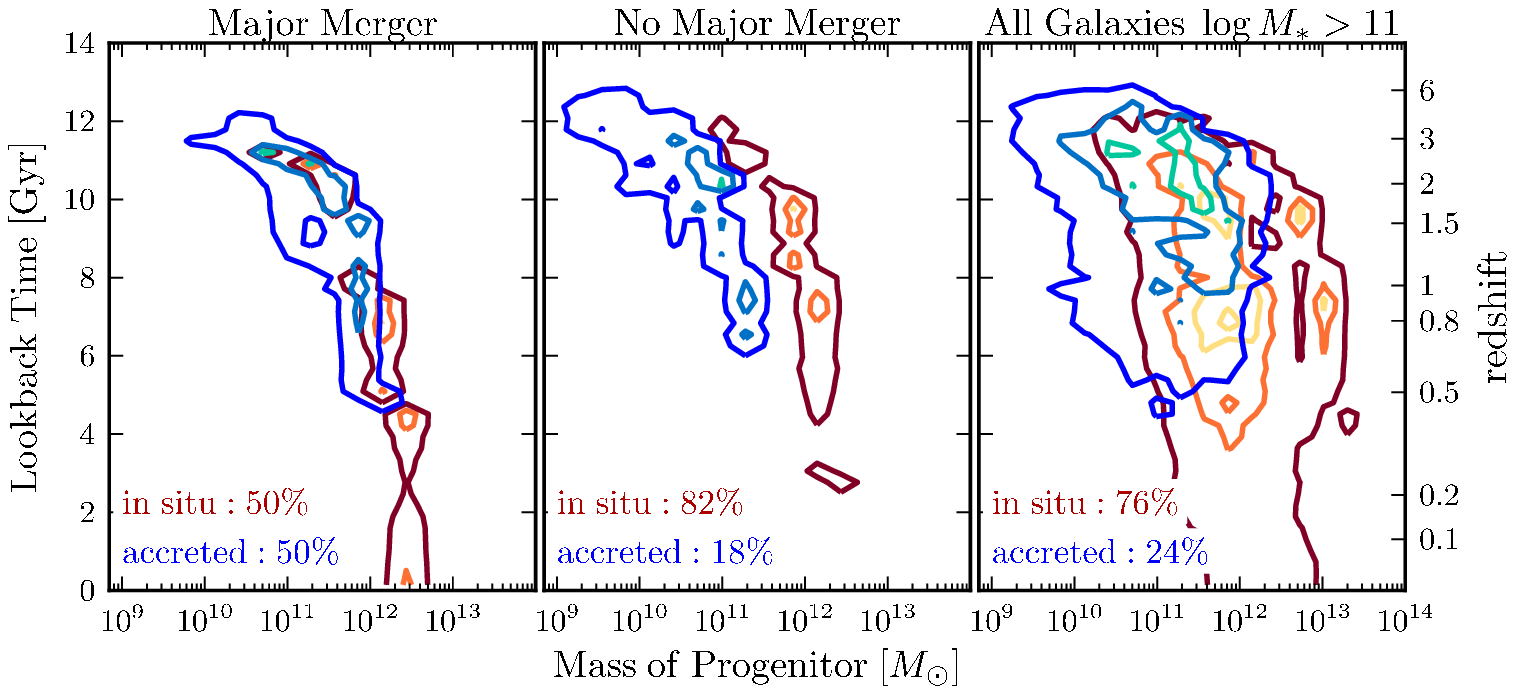}
    \caption{Distributions of stellar ages and host galaxy 
    masses for
      accreted (blue contours) and in-situ (red contours) stars. The
      x-axis shows the mass of the galaxy in which a given star
      particle formed. The left panel shows the distribution for a
      galaxy which underwent a major merger at $z\approx0.4$. The
      center panel shows a galaxy which had no major mergers. The
      right panel shows the distribution of stars for all massive
      galaxies in the simulation. The contours have been smoothed and
      enclose $20$, $50$, and $80$ per cent of the stars. {\rm The
        differences between the accreted and in-situ stars are largest
      for galaxies with no major mergers. }}
    \label{fig:mhostz}
  \end{figure*}
}

\ifthenelse{\boolean{astroph} \or \boolean{mn2e}}{
  
  \newcommand{\tailfig}[1]{{}}
}

\title[Accretion vs. in-situ star formation]{Building galaxies by accretion and in-situ star formation}
\author[C.~N.~Lackner]{C.~N.~Lackner$^{1}$\thanks{E-mail:
    clackner@astro.princeton.edu}, R.~Cen$^1$, J.~P.~Ostriker$^1$, and
  M.~R.~Joung$^{2,3}$ \\
$^{1}$Department of Astrophysical Sciences, Princeton University, Princeton, NJ
08544\\
$^{2}$Department of Astronomy, Columbia University, 550 West 120th Street, New York, NY 10027\\
$^{3}$Department of Astrophysics, American Museum of Natural History, 79th Street at Central Park West, New York, NY 10024}
\begin{document}
\date{\today}
\pagerange{\pageref{firstpage}--\pageref{lastpage}} \pubyear{2011}
\maketitle

\label{firstpage}

\begin{abstract}
We examine galaxy formation in a cosmological AMR simulation, which
includes two high resolution boxes, one centered on a $3\times
10^{14}\ \Msun$ cluster, 
and one centered on a void. We examine the
evolution of $611$ massive ($M_* > 10^{10} \Msun$) galaxies. We
find that the fraction of the final stellar mass which is  
accreted from other galaxies is between $15$ and $40\%$ and increases
with stellar mass. The accreted fraction does not depend strongly on
environment at a given stellar mass, but the galaxies in groups and
cluster environments are older and underwent mergers earlier than
galaxies in lower density environments. On average, the accreted stars
are $\sim 2.5$ Gyrs older, and $\sim 0.15$ dex more metal poor than
the stars formed in-situ. Accreted stellar material typically lies on
the outskirts of galaxies; the average half-light radius of the
accreted stars is {\rm $2.6$ times larger} than that of the in-situ
stars. This leads to radial gradients in age and metallicity for
massive galaxies, in qualitative agreement with observations. {\rm
  Massive galaxies 
grow by mergers at a rate of approximately $2.6\%/ \mathrm{Gyr^{-1}}$}. These
mergers have a median (mass-weighted) mass ratio less than $0.26\pm
0.21$, with an absolute lower limit of $0.20$, for galaxies with
$M_*\sim10^{12}\ \Msun$. This suggests that major mergers do not
dominate in the accretion history of massive galaxies. All of these
results agree qualitatively with results from SPH simulations by
\citet{Oser10, Oser11}.
\end{abstract}


%
\section{Introduction}
\label{intro}

In a \LCDM{} cosmology, galaxy growth is hierarchical; large galaxies
form by accreting smaller systems of gas, stars, and dark matter. This
model has been extensively tested by simulations and
observations. However, it is not known whether stars or gas
dominate the accretion history of galaxies, and what 
effects 
this accretion has on the observable properties of the galaxies. With
modern hydrodynamical simulations, we can directly measure what
fraction of the stars in a massive galaxy have been 
formed in-situ and
what fraction have been made in other galaxies which were subsequently
accreted by the parent galaxy. Similarly, we can map out when
accretion occurs, what kind of accretion events dominate, and how the
accretion affects the observable properties of galaxies. In this work,
we explore these effects for a set of massive galaxies from a
hydrodynamical cosmological simulation. 

\citet{Oser10} (hereafter Os10) present a useful framework for studying the
hierarchical build-up of galaxies. {\rm Based on hydrodynamical simulations
of early-type galaxy formation, they suggest that massive galaxy
formation can be divided into two phases: an early, rapid in-situ star
formation period followed by a late merger-dominated period. The
former period is similar to the picture previously called
``monolithic collapse'' \citep{ELS62, Larson1975, Carlberg1984} and
{\rm currently} labelled ``cold
flow-driven'' star formation
\citep[e.g.][]{Dekel06,Dekel09,Elmegreen09, Ceverino2010}, while the
latter phase is similar to prior pictures of galactic cannibalism
\citep{OstrikerHausman77, White1978, SearleZinn78, Kauffmann1993}.} In
this scheme, individual stars can 
be classified according to whether they formed within the virial
radius of the final system (in-situ) or 
outside of it (accreted). Os10 show that this distinction between stars is
unambiguous; there is a large spatial separation between the birth
places of stars formed in-situ and those added in accretion
events. They further show that massive galaxies are dominated by
accreted stars, which are typically older than stars formed
in-situ. In this work, we find that the accreted stars in a galaxy are
older, more metal-poor, and found at larger radii in the final
galaxy than stars formed in-situ. The metallicity trends 
are 
consistent with observations of
radial trends in colour \citep[e.g.][]{deVaucouleurs61,   Tortora2010,
  Tal2011}, metallicity \citep{Spinrad1971, Faber1977,   Davies1993,
  Brough2007,   Rawle2008,   Spolaor2010, Kuntschner2010,
  Coccato2011} and globular cluster metallicities \citep{Forbes2011,
  Arnold2011} in early-type galaxies.

{\rm The differences between accreted and in-situ stars are also
  seen in disc galaxies. For example, in the Milky Way stellar halo,
  there has long been observational evidence 
  for at least two stellar populations. The inner halo
  consists of stars with high $\alpha-$element abundances
  \citep[e.g.][]{Nissen2010} and rotates with the galactic disk
  \citep{Deason2011}, while the outer halo has low $\alpha$-element
  abundances and is not rotating, suggesting it was created by
  accretion of satellite galaxies. Simulations of disc galaxies show
  that the differences in dynamics and chemical abundances in the
  stellar halo are dues to the differences in the accreted and in-situ
  stellar populations \citep{Brook2004, Abadi2006, Zolotov2009,
    Zolotov2010, Font2011}. }

The two phase model of galaxy formation 
(i.e., early,  
in-situ star formation and late accretion) is also useful in
explaining the evolution of compact, massive ellipticals at $z \sim 2$
\citep{Trujillo07, vanDokkum08, Damjanov09, vanDokkum2010,
  Oser11}. Both direct profile measurements and velocity dispersions
have shown that these systems are $\sim 100$ times denser (within $1$
effective radius) than present-day ellipticals of the same mass
\citep{Daddi2005, vanDokkum08, vanderWel2008, vanderWel2011,
  vandeSande11}. However, if compared on same {\it physical} scale,
the densities in the central portions of $z\approx2$ and present-day ellipticals are similar,
suggesting that early-type galaxies have increased their size through
minor, dry mergers which add stars to the outskirts of massive
galaxies \citep[][but see \citealp{Newman2012}]{Naab07, Naab09,
  Bezanson09, Hopkins09a, Carrasco2010, Oser11, Tal2012}. {\rm By using
close companions to estimate the merger rate of compact galaxies at
$0.4<z<2.5$, \citet{Newman2012} observe that minor merging may not be
sufficient to explain the size evolution of compact galaxies from
$z\approx2$.} However, the size evolution due to minor mergers has
been found in hydrodynamical simulations \citep[e.g.][]{Naab09,
  Oser11} and is consistent with the size growth found by observations.

In this work, we make use of galaxy catalogs from cosmological
simulations done with adaptive mesh refinement (AMR)
\citep{Cen2010}. {\rm The use of an Eulerian grid-based code instead of a
Lagrangian particle-based code is a notable distinction between this work and that
of Os10. \citet{Scannapieco2011} find that different numerical
hydrodynamics can yield differences of a factor of two in simulated
galaxy properties. However, \citet{Scannapieco2011} find that changes
in the feedback implementation can yield even larger differences in
the simulation results. We will address the differences in sub-grid
physics between our work and that of Os10 below. }
The AMR 
simulation studied here contains two high resolution boxes, one centered on a
galaxy cluster and another centered on a void. 
These large volumes
simulated at high resolutions 
allow 
us to study the merging histories
of more than $600$ galaxies in a variety of environments, from void to
group to cluster. Following Os10, we have divided the stars in each galaxy
into accreted and in-situ, which allows us to examine the star
formation histories for the galaxies as a function of stellar mass and
environment. The qualitative similarities between our results and those of
Os10, despite profound differences in numerical techniques and sub-grid
physics, help substantiate the two-phase model of galaxy evolution for
early type galaxies.

The paper is divided into the following sections:
\S\ref{sec:explainSim} details the simulation, the building of the
merging histories and the tagging of stars as accreted or in-situ;
\S\ref{sec:insitu_acc} and \S\ref{ssec:compareOser} describe the
properties of the in-situ and accreted stars and compare our
simulation results to those of Os10. \S\ref{sec:mergers} focuses on
the merger histories and the types (major 
vs. 
minor) of mergers the
galaxies in our simulation underwent and their effects on the final
galaxy properties. For the
simulation and throughout this work, we use the following \LCDM{} 
cosmology, consistent with WMAP-7 \citep{Komatsu11}:  $\Omega_M =
0.28$, $\Omega_\lambda = 0.72$, $\Omega_b = 0.046$, $\sigma_8 = 0.82$,
$n=92$, $H_0 = 100\h\kmps \Mpc^{-1}$, and $h=0.7$.

\section{Description of Simulation}
\label{sec:explainSim}

This work uses the same cosmological simulations as
\citet{Cen2010,Cen2011}, and further 
descriptions of the methodology can be found
there. The simulations are performed with the AMR
Eulerian hydrodynamics code, Enzo
\citep{Bryan99,OShea04,Joung09}. {\rm Enzo uses the particle-mesh
  technique for the dark matter N-body dynamics. Poisson's equation
is solved using a fast-Fourier transform on the base grid and
multigrid relaxation on the the higher resolution subgrids
\citep{OShea04}. Enzo solves the hydrodynamics using a modified version
of the piecewise parabolic method \citep{Woodward1984,OShea04}, which
is formally second-order accurate and conserves energy, momentum and
mass fluxes. }For this work, two regions, a cluster and a void, were selected from a low resolution simulation of a periodic box 120 $\h^{-1}$Mpc on a
side. These regions were re-simulated separately at high resolution. The cluster box is
$21\times 24\times 20\ \h^{-3}$Mpc$^3$, and the void box is
$31\times 31\times 35\ \h^{-3}$Mpc$^3$. These regions represent 
$+1.8\sigma$ and $-1.0\sigma$ fluctuations, respectively. The refined
regions are surrounded by two buffer layers, $\sim1\h^{-1}$Mpc thick,
which connect to the underlying low-resolution periodic simulation
box of $120 \h^{-1}$Mpc. The resolution in the refined regions is $460\
\h^{-1}\mathrm{pc}$ physical or better. {\rm The two refined regions
  are similar to two of the five 
  regions taken from the Millennium Simulation \citep{Springel2005} and then
  simulated at high resolution in \citet{Crain2009}. 
 The volume of the void box used here is $5$ times larger than the
 regions simulated in \citet{Crain2009}, while the cluster box is
 comparable in size to the $+2\sigma$ region in \citet{Crain2009}. }

In the refined regions, the dark matter particle mass is
$1.07\times 10^{8}\ \Msun$, and the stellar particle mass is typically
$10^6\ \Msun$. Each stellar particle is tagged with its initial mass,
metallicity, dynamical time, and formation time. {\rm Additionally, the
luminosity of each star particle in the five Sloan Digital Sky Survey
(SDSS) filters \citep{Fukugita1996} is computed using the Galaxy
Isochrone Synthesis Spectral Evolution Library (GISSEL)
stellar synthesis code \citep{Bruzual2003}, assuming a Salpeter initial
mass function (IMF).  

The
simulations include a UV background \citep{Haardt96}, a prescription
for shielding from UV by neutral hydrogen \citep{Cen05}, and
metallicity dependent radiative cooling \citep{Cen95}. The UV
background radiation field is subject to self-shielding; we compute
the local ionization and cooling/heating balance 
for each cell in the following way.
A local optical depth approximation is adopted to crudely mimic
the local shielding effects: each cubic cell is flagged with six
hydrogen ``optical depths'' on the six faces, each equal to the product of
the neutral hydrogen density, hydrogen ionization cross section and
scale height. The appropriate mean from the six values is then
calculated. Equivalent means for neutral helium and singly-ionized helium are also computed.
Metal cooling is computed with ionization equilibrium including the effects
of the redshift-dependent UV-X-ray background, using a code based on
the standard Raymond-Smith code \citep{Raymond1976}.
The cosmological reionization in this model occurs at redshift
$z=8-9$.}

 Star particles
are created in cells that satisfy the criteria set forth in
\citet{Cen92}. Specifically, stars are created when the baryon
overdensity, 
$\delta\rho_b/\rho > 5.5$ and the following are true:
\begin{subequations}
  \label{eq:sfr_crit}
  \begin{align}
    \nabla \cdot  \mathbf{v} &< 0\quad , \\
    t_{\mathrm{cool}} &< t_{\mathrm{dyn}} \equiv
    \sqrt{\frac{3\pi}{32\mathrm{G}\rho_{\mathrm{tot}}}}\quad ,\ \mathrm{and}\\ 
    m_b &> m_{\mathrm{Jeans}} \equiv
    \frac{c_s^3}{\mathrm{G}^{3/2}\rho_b^{1/2}(1+\rho_b/\rho_d)^{3/2}}\quad ,
\end{align}
\end{subequations}
where $c_s$ is the sound speed and 
$\rho_b$ and $\rho_d$ 
are the baryon and dark
matter densities, 
respectively. 

Supernovae feedback is modeled as in 
\citet{Cen05}. {\rm Feedback thermal energy and metal-enriched ejecta is
  distributed into the $27$ cells surrounding the star
  particle. The amount added to each cell is inversely proportional to
the density of each cell, mimicking the process of supernovae blast
waves, which channel energy into the least dense surrounding
regions. The rate of feedback has the form
$1/t_{\mathrm{dyn}}\left[(t-t_i)/t_{\mathrm{dyn}}\right]\exp\left[-(t-t_i)/t_{\mathrm{dyn}}\right]$,
 where $t_i$ is the formation time of a given star particle, and
 $t_{\mathrm{dyn}}$ the dynamical time of the cell. Accounting for both Type II and prompt Type I supernovae,
the total amount explosion kinetic energy per stellar mass ($M_*$)
formed (assuming a Chabrier IMF) is $e_{\mathrm{SN}}M_*c^2$, where
$e_{\mathrm{SN}}=1\times10^{-5}$. The feedback process is entirely
hydrodynamically coupled to the surroundings and subject to all
physical processes, such as cooling.  The simulations do not include
input from active galactic nuclei; its absence may contribute to the  excess of
massive galaxies.  

Several tests of the feedback
process for these simulation are presented in earlier
works. \citet{Cen2010} show that the kinematic properties of  damped
Lyman-alpha systems in the simulations agree well with observations,
providing an significant test to the feedback model. In
\citet[][]{Cen2011} we show that the phenomenon of cosmic downsizing
can be naturally explained in our model with the necessary
large-volume cosmological setting, due to correlations between galaxy
environment and astrophysical processes that operate on gas physics. In
\citet[][]{Cen2011c} we show that galaxy luminosity functions for both
UV and FIR selected galaxies can be self-consistently produced by the
simulation, when reprocessing of optical-UV radiation into IR
radiation is followed in detail. In \citet[][]{Cen2011d}, we show that
our model reproduces a variety of observed properties of O~VI $\lambda
\lambda$1032, 1038 absorption lines at $z\sim 0$. In combination,
these tests strongly indicate a range of applicability of our
simulations to complex systems, from galaxies at kpc scales to Mpc
scales in the intergalactic medium, and empirically validate our star
formation feedback prescription, upon which all these phenomena
critically depend. This validation of the simulation results is
critical and allows us, with significant confidence, to perform the
particular analysis presented here.}  

\subsection{Merger trees}
\label{ssec:mergertrees}

In order to identify galaxies in a given snapshot, we use the
HOP algorithm on stellar particles
\citep{Eisenstein99,Cen2011}. {\rm For the HOP program we choose an outer
stellar overdensity of $\delta_{outer}=10000$, and density ratios of
$\delta_{saddle}=2.5\delta_{outer}$ and 
$\delta_{peak}=3\delta_{outer}$; these ratios are recommended by
\citet{Eisenstein99}.} We have examined $600$ groupings with $M_* >
5\times 10^9 \Msun$ at $z=0$ in the cluster simulation and have found
$9$ HOP groupings which contain more than one (usually two) visually
distinguishable 
galaxies. The central galaxy of the cluster is included in one such
grouping.  We find no such groupings in the void box. This represents
a contamination of $\la 2\%$ of our galaxy sample. For the remainder
of the paper, we will 
exclude the $9$ groupings which have not been adequately separated by
the the HOP algorithm at $z=0$. 

In order to study the accretion histories of the galaxies, we build
merger trees for the simulations. For the cluster box, we use $38$
snapshots: one every $\Delta z = 0.05$ between $z=0$ and
$1.35$. Beyond $z=1.35$, we have snapshots at $z=1.5$, $1.6$, $1.75$,
$1.9$, $2.0$,  
$2.2$, $2.5$, $2.8$, $3.1$, and $4$. For the void box, we only use
$14$ redshift slices at $z=0$, $0.05$, $0.15$, $0.2$, $0.4$, $0.5$,
$0.6$, $0.8$, $1.0$, $1.6$, $1.9$, $2.5$, $3.1$ and $4$. The merger
trees are built by looking at the overlap of the stellar particles in
galaxies in sequential redshift slices. For example, if most of the
stars in galaxy B at $z=0.05$ are later in galaxy A at $z=0$, we
label galaxy B a parent of galaxy A. Galaxy A may have many parents at
$z=0.05$; we call the main progenitor (the same galaxy at an earlier
time) the galaxy that contributes the most mass to the child
galaxy. The remainder of the parent galaxies are galaxies which have
merged into the main galaxy during the current redshift step. This
process can be repeated for all the galaxies in all the redshift
slices. Each galaxy has a progenitor and possibly a set of galaxies
which have just merged into it. Stepping back through the main
progenitors gives a full history for a given galaxy. 

We verify this method of grouping by using every other redshift slice
and 
checking that the progenitors do not change. Ninety-two percent
($91\%$) of the galaxy merger trees in the cluster (void) box are the same
to $z=1$ if we only use every other redshift slice. The galaxies with
modified histories are typically 
below our mass threshold. {\rm We also check that
the $200$ most-central particles in a galaxy have significant overlap
with the $200$ most-central particles in its main
progenitor. This is true for more than $98\%$ of the galaxies with
masses above $5\times10^9$ for the cluster box and $95\%$ of the
galaxies in the void box. The majority of galaxies which fail this
test cannot be traced back to $z > 2.0$ and will not be used in our analysis. }

After identifying merger histories for each galaxy, we can label the
stars in every $z=0$ galaxy as accreted or in-situ. As above, we proceed
in a stepwise manner toward high redshift snapshots. For each stellar
particle in a galaxy at a given redshift, we examine where it was in
the previous redshift slices. Star particles which were in the main
progenitor galaxy are labelled as in-situ. Stellar particles in any of
the other parent galaxies are labelled as accreted, {\rm and the accretion
redshift is noted}. The remainder of stellar particles were either
formed during the current redshift step, and labeled as in-situ, or
they formed earlier and are labeled as accreted. These stellar
particles are due to accretion from unresolved galaxies, as well as
uncertainty in the grouping algorithm. If a star particle was born in
the main progenitor galaxy, but later 
re-accreted, we consider it an in-situ star particle. Such particles
contribute between $2$ and $5$ per cent to the total galaxy stellar
mass. We then examine the stars of the main progenitor (those which
were initially labeled in-situ) in the same way and repeat the process
for all higher redshift slices. This allows us to distinguish between
accreted and in-situ stellar material to our highest redshift snapshot
at $z=4$.  

One difficulty in this method is ensuring that none of the stars we
consider in-situ were actually 
formed outside the galaxy and accreted
between snapshots. Os10 find a large
separation in
space ($3$ orders of magnitude in distance from the centre of the main
progenitor galaxy) between in-situ and accreted 
particles. However, their Figure 7 also shows the birth places of a
population of accreted stars which formed between $0.1\
r_{\mathrm{virial}}$ and $1.0\ r_{\mathrm{virial}}$. If our redshift
snapshots are not sufficiently close together, we will count these
stellar particles as in-situ, instead of accreted. We have compared
the dynamical time of a virialized halo ($t_{dyn} \sim 1/\sqrt{G\
  200\rho_{\mathrm{crit}}} \sim 1/(10 H(z))$), to the time between
redshift slices. For the cluster, the dynamical time of 
the halo is larger than the time between redshift slices by at least a
factor of $2$ for $z \le 1.35$. Beyond $z=1.35$, the halo dynamical
time is comparable to the density of redshift
snapshots. Therefore, our calculations of the accreted fraction may
not include stars formed during infall to the final galaxy, and should
be considered lower limits.

\subsection{Galaxy population}
\label{ssec:galaxyPop}
Before examining the merger histories of the simulated galaxies, we
have examined the overall properties of the galaxies at
$z=0$. \citet{Cen2011} have shown that the luminosity function for
galaxies in the cluster and void agree well with SDSS observations \citep{Blanton03b}, except at the highest
luminosities, where the simulation produces too many luminous
galaxies. This is due to insufficient feedback; the simulations do not
include AGN feedback or stellar feedback from Type I
supernovae. Additionally, the highest mass galaxies suffer from
overmerging \citep{White76, Moore96}. These highest luminosity bins
can be corrected by taking into account AGN feedback and decreasing
the stellar mass by a factor of $1/(1+(M_{\mathrm{halo}}/1 \times
10^{13}\ \Msun)^{2/3})$ \citep{Cen2011,Croton06}.

Figure
\ref{fig:massFunc} shows the uncorrected mass function of galaxies in
each region at $z=0$,
compared to the mass function of SDSS galaxies \citep{Li2009}.  We have only
included galaxies with at least $100$ dark matter particles
($M_{\mathrm{halo}} \gtrsim 10^{10}\Msun$). {\rm The {\it average} mass
function for the simulation is a weighted sum of the cluster and void
box. Using the same simulations, \citet{Cen2011} suggest a weighting
of $6:1$ for the void and cluster box, based on matching the simulated
star formation rate history to observations. The weighting
scheme of \citet{Crain2009}, yields a weighting of $2:3$ for the void
and cluster box, since the cluster box is $63\%$ smaller than the void box,
but it's also a rarer density perturbation. For the remainder of this work,
we weight the galaxies in the void and cluster box equally, unless
otherwise noted.}

There is a clear deficit of galaxies with masses
below $\sim3\times 10^{10}\ \Msun$. In fact, if we include galaxies
with fewer dark matter particles, both mass functions turn up again and have
second low mass peaks at $M_* \sim 10^8\ \Msun$, however, these groupings
are most likely due to resolution effects. The simulated mass functions bracket
the observations well in the range 
$10.5 \lesssim \log M_*/\Msun \lesssim 11$, 
but
as explained above, the simulation overproduces massive
galaxies. {\rm The overproduction of massive galaxies is a typical
  problem for hydrodynamical simulations and is due to insufficient
  feedback to suppress star formation and to overmerging among close
  pairs of galaxies. This is a serious limitation of 
  our work, and the numerical values we derive from the simulations
  will certainly change with higher resolution, improved feedback
  models, and lower star 
  formation rates. 

Since improved feedback will reduce the stellar masses of
  all massive galaxies, it is not obvious if it will decrease number
  of accreted or in-situ stars more. The high levels of star
  formation at $z=0$ (see \S\ref{ssec:compareOser}) suggests that most
  of the excess stars are formed in-situ, but only improved
  simulations will give a definitive answer. Nonetheless, we
  are confident in our results for two main reasons. The first is that
  the differences between accreted and in-situ stars do not depend strongly
  on mass (see Figures
  \ref{fig:mGroup_ages}--\ref{fig:mGroup_profile}), and are present even
  in the lowest mass bin where the mass function of galaxies is
  reasonable. Secondly, although improved feedback will certainly
  modify the mass function, it will not substantially alter the
  rank-ordering of 
  galaxies; the most massive galaxies in our simulation will still be
  the most massive. Therefore, we expect that trends with stellar mass
 will have different slopes, but the sign of those slopes should be
 insensitive to the feedback model. 
} In
the next sections, we will only examine the properties of galaxies
with at least $100$ dark matter particles and stellar masses larger
than $3\times 10^{10}\Msun$, {\rm but we continue to include the most
  massive galaxies despite their overabundance}.   
\figMfunc

\subsubsection{Galaxy formation efficiency}
\label{sssec:galformEffic}
Simulations from cosmological initial conditions are known to
overproduce stars \citep[][and references therein]{Oser10,
  Guo2010}. {\rm This is true for the simulations in this work as well,
  and can be easily shown by comparing the mass ratio of dark matter
  and stars to observations. The ratios of stellar mass to dark matter
  mass 
  at $z=0$ are $1.2\times10^{-2}$ and $6.0\times10^{-2}$ for the void
  and cluster boxes, respectively. The two weighting schemes
  discussed above give an average ratio in the range $1.9$ to
  $4.1\times10^{-2}$ for the entire simulation. The
  observed ratio in the local universe is $(1.2\pm0.3)\times10^{-2}$
  \citep[e.g.][]{Fukugita2004,Gallazzi2008}. The void box has a star
  formation efficiency which agrees well with the observed average. However,
  observations show that underdensities in the universe are also
  under-luminous \citep{Peebles2001}. Therefore, if the star formation
  and feedback in the simulation were accurate, the void box would
  have a lower star to dark matter ratio than the average. Thus,
  stars are overproduced in both the cluster and the void box, with
  the overall star formation rate too high by a factor of two to four. }

We can examine the overproduction of stars in the cluster box in more
detail by computing the ratio $f_* =
M_*/M_{\mathrm{DM}}\times (\Omega_{\mathrm{DM}}/\Omega_b)$ for both
simulated and observed galaxies and their host dark
matter halos. In the simulations used here, the central cluster has a virial radius (using the radius inside which
$\langle \rho \rangle = 200 \rho_{\mathrm{crit}}$) of $1.3\ \Mpch$. The
total dark matter mass interior to this radius is
$3.0\times10^{14}\ \Msun$, 
and the total stellar mass interior to this
radius is $3.0\times10^{13}\ \Msun$. Therefore, the efficiency of star
formation in the cluster is $f_* = 0.60$, or $60\%$ of the baryons in
the cluster have been turned into stars by $z=0$. For a cluster of
this mass, the expected star formation efficiency from weak lensing
and halo occupation distribution (HOD) methods is $10\% \la f_* \la
15\%$ \citep{Leauthaud2012}. This range is slightly above the values
found by matching simulated DM halos to the observed galaxy mass
functions \citep{Guo2010, Behroozi2010, Moster2010}. Thus, the star
formation in the cluster is over-efficient by roughly a factor of {\rm
  four}. The simulation star formation efficiency is in part too high
due to the lack of sufficient feedback from supernovae and AGN. If we
correct the stellar masses to take this into account, as in the
previous section, the efficiency decreases to $\sim30\%$. 
\figaccisages
\figaccismetals

\subsubsection{Mean stellar properties}
The top panel in Figure \ref{fig:m_agesISACC} shows the relation
between galaxy stellar mass and SDSS $r-$band luminosity-weighted
age. The simulated trend is compared the observed mass-age relation
from SDSS (thick gray line) \citep{Gallazzi2005}. Although there is
little trend with mass, the youngest galaxies are among the least
massive. The lack of trend at high masses is consistent with
observations from SDSS \citep{Gallazzi2005}, in which 
the steepest portion of the stellar mass--stellar age trend occurs in
the mass 
range $3\times 10^9 \la M_* \la 3\times 10^{10}$. However, the median
ages for the galaxies in our sample are several Gyrs younger than
observed galaxy ages \citep{Gallazzi2005}. This is partially due to an
excess of star formation at late times (see Figure
\ref{fig:mGroup_ages}) which is related to the insufficient resolution
and feedback
modeling in the simulation. Figure \ref{fig:m_agesISACC} also shows that
the median age of galaxies is dependent on environment, with the
oldest galaxies found in the cluster box, while none of the void box
galaxies are older than $\sim 7\ \mathrm{Gyr}$. Thus the trends of
galaxy age with stellar mass and local density are {\rm broadly}
consistent with the observational picture of downsizing and the
quenching of star formation in cluster environments {\rm
  \citep[see][for a study of downsizing in this simulation]{Cen2011}.}  

It is expected that the trends in
metallicity will be similar to those for stellar ages. As with the
stellar ages, we find that stellar 
mass
and stellar metallicity
are not strongly correlated in the simulations (see Figure
\ref{fig:m_zsolISACC}, top panel). There is a stronger dependence
on metallicity with environment; the void box galaxies have 
metallicities around $Z\approx 1\ Z_{\odot}$, while the cluster galaxy
metallicities are centered at $Z \approx 1.5\ Z_\odot$ with very
little scatter. This value of metallicity is consistent with the
median metallicity for the most massive ($M \ga 10^{11}\ \Msun$)
nearby SDSS galaxies \citep{Gallazzi2005} (grey solid line).

{\rm The excess of young stars and overabundance of metals suggests
  again that the simulations suffer from over-cooling and insufficient
  feedback. \citet{Wiersma2011} find that changes in the feedback
  prescription can alter the stellar metallicity by factors of
  $4$. Their simulations suggest that the inclusion of AGN feedback
  and momentum-driven supernovae winds can bring the stellar
  metallicity into agreement with observations.}

\section{In-situ vs.  accreted stars}
\label{sec:insitu_acc}
\figaccrfrac
In order to
examine the accretion history of galaxies, we restrict our sample to
galaxies with robust merger histories. We require that galaxies
have a stellar mass $M_* > 10^{10} \Msun$, and at least $100$ dark
matter particles. We also require either that a galaxy can be traced
back to $z > 2.0$ or that its smallest progenitor of the galaxy has a
stellar mass $M_* < 10^9 \Msun$. This removes galaxies which have gaps
in their merger history, but retains galaxies which have grown
significantly since $z=2$. The sample includes $443$ galaxies from the
cluster box and $168$ galaxies from the void box. These represent
$\sim 85\%$ of the $z=0$ galaxies which 
satisfy the stellar and dark matter mass 
constraints. {\rm We test that this smaller sample is not a biased
  subset of the larger sample using the Kolmogorov--Smirnov (KS)
  test. We find the probabilities that the distributions of
  stellar mass, stellar age, and accreted fraction are the same for
  both samples are $1.00$, $0.99$, and $0.98$, respectively. }

The top panel in
Figure \ref{fig:m_macc} shows the fraction of the final stellar mass
which was accreted as a function of stellar mass. This figure (and all
subsequent figures) only include galaxies for which we have robust
merger histories. We find that the
accreted fraction increases as a function of stellar mass from
$17\pm7\%$ for $10^{10} \la M_* \la 10^{11}\Msun$ to $\sim40\pm15\%$
for the most massive galaxies in the cluster box ($M_* \ga
10^{12}\Msun$). This value is in agreement with estimates of the
merged stellar mass for massive cluster galaxies from observations of
paired galaxies \citep{Lin2010}. Although the sign of this trend is consistent with 
that found in Os10, both the slope and the zero-point are
significantly smaller in our sample. The causes of this discrepancy will be
discussed further in \S\ref{ssec:compareOser}. {\rm We find no trend with
the dynamical age of system and the accreted fraction, in contrast to
the results of \citet{Font2011} and \citet{Zolotov2009}. The excess of
late in-situ star formation in our simulation may modify the dynamical
ages of the galaxies and alter any trend in accreted fraction with
age. }

{\rm The trend of accreted fraction with stellar mass is sensitive to
  the excess of massive galaxies in the simulations. If we restrict
  the sample to galaxies with $M_* < 10^{11}$, the slope of accreted
  fraction with mass is consistent with zero. Furthermore, overmerging
  will modify the slope of this relation. If the {\it real} relation
  between $\log M_*$ and the accreted mass fraction, has a small or
  negative slope, then additional 
  merging will tend to increase the slope; if the real slope is very
  steep, then overmerging will tend to flatten the slope. The value
  for which the slope remains unchanged is $\sim0.09$, and depends only
  weakly on the stellar mass and average merger ratio. This value is
  close to the slope we measure slope, and suggests the
  slope in Figure \ref{fig:m_macc} may be an artifact of
  overmerging, and the real slope could be either shallower or steeper.
}

Direct comparisons of the merged fraction with observations is
difficult. Although there are many studies of the merger rate as a
function of redshift \citep[e.g.][see \citealp{Lotz2011} and
references therein]{LeFevre00,Bundy2004,Lin2004,Kartaltepe2007,Lotz2008,Jogee2009,Bridge2010}, these studies suffer from many
observational biases, including difficulty in selecting comparable
descendant/progenitor populations. Nonetheless, studies of
paired galaxies indicate a weak positive trend in the major merger rate with
progenitor galaxy mass \citep[but see
\citealp{Bridge2010}]{Bundy2009,Darg2010}. This is compatible with our
results of an increase in the accreted fraction with stellar mass,
provided that the average merger ratio does not decrease precipitously
with galaxy mass.

\subsection{Comparison to Oser, et al. 2010}
\label{ssec:compareOser}
The upper and lower panels in Figure \ref{fig:m_macc} show the
accreted fraction of stars for the galaxies in this work and in Os10,
respectively. Both studies find the same sign for the relationship
between accreted fraction and total stellar mass, but Os10 find an
accreted fraction about twice as large as that found in this
work. Furthermore, the slope of the Os10 relation is about twice as
steep (although the mass range of their sample is smaller). We suggest
that these differences in accreted fraction can be explained by {\rm several
major differences in the simulations in the resolutions, the feedback
models, and the star formation efficiencies.}

As shown in Figure \ref{fig:mGroup_ages}, even the most massive
galaxies in our work experience significant late star
formation. The median $z=0$ SFR for galaxies with stellar masses above
$10^{11}\ \Msun$ is $5.2\ \Msunpyr$, above the observed star formation
rate for most present-day massive galaxies. This
ongoing star formation contributes significantly to the in-situ
fraction of stars. \citet{Naab07} have shown that 
higher resolution can decrease the amount of late time star formation (see their Figure
2). For the lower resolution simulations, the $z=0$ SFR is $\sim
5\ \Msunpyr$. For smoothed particle hydrodynamics (SPH) codes, like the
GADGET code used in \citet{Naab07} and Os10, the low resolution suppresses the
Kelvin-Helmholtz instability 
\citep{Agertz07}, which breaks up clumps of cold
gas before they sink to the center of a galaxy and form in-situ
stars. {\rm At higher resolutions ($m_* < 10^5\ \Msun$), the final
  star formation rates is less than $1\ \Msun\mathrm{/yr}$, even without
  AGN or Type I supernovae feedback \citep{Johansson2012}.}

Although the AMR code used in this work does
not suffer from the same problem as SPH codes at low resolution, low
resolution  AMR 
simulations will suppress
fragmentation of infalling gas clouds. These gas clouds will survive
to the center of the nascent galaxy, where they will form stars
in-situ at late times. The high resolution simulations of Os10 use
particle masses of 
$m_{*,\ \mathrm{gas}} = 6.0\times 10^6 \Msun$ and $m_{\mathrm{dark}} = 3.6\times 10^7\Msun$. 
Although not the highest
mass resolution tested in \citet{Naab07}, this particle resolution is
large enough to suppress most spurious late star 
formation. Furthermore, the simulation in Os10 have a factor 
$3$ more
mass resolution in dark matter than the simulations examined in this
work. Thus, we expect additional late in-situ star formation in our
simulations, and, therefore, a lower accreted fraction when compared to
the simulations of Os10. {\rm \citet{Cen2010} have performed spatial
resolution tests for the cluster box and find that the overall
convergence is quite good. However, they find that the metallicity in
damped Lyman-alpha systems increases by around $0.2$ dex for an AMR
simulation with spatial resolution better by a factor of
two. Therefore, we
expect that 
improved spatial resolution will alter the star formation and feedback
processes, changing the the in-situ and accreted fractions. The same
is certainly true for improved mass resolution, which will be tested
in future work.}

{\rm Both the simulations presented here and those in Os10
  suffer from an overabundance of star formation, and a deficit
  of feedback. The differences in the feedback and cooling models may
  help explain the difference in accreted fraction between the two
  simulations. \citet{Scannapieco2011} show that the addition of
  metal-line cooling to \verb+GADGET3+ (used by Os10) can help
  decrease the final stellar mass of galaxies by $\sim 30\%$, but does
  little to the star formation history. Therefore, we expect
  additional cooling will have little effect on the fraction of
  accreted stars. However, \citet{Scannapieco2011} show that
  different feedback models can greatly affect the star formation
  history in simulations. For various models of feedback and
  cooling in SPH  simulations, \citet{Scannapieco2011} find that the
  redshift by which half of the stars 
  have formed ranges from $\sim1.5-4.0$, comparable to the redshift
  range of the median stellar age from Os10. Additional supernovae
  feedback, especially the addition of supernovae winds, tends to
  decrease the median redshift of star 
  formation. We expect that decreased early star formation will
  decrease the accreted fraction, as fewer small systems will form
  stars at high redshift to be accreted later. 

In the work presented
  here, star formation occurs much later than in the work of Os10;
  Figure \ref{fig:mGroup_ages}  shows half the stars in our
  simulations are only formed by $z\approx1$, in agreement with results from other AMR codes
  \citep{Scannapieco2011}. As mentioned above, late star formation
  will lead to an increased in-situ fraction in massive galaxies. We expect 
that additional feedback will decrease star formation at late times
and shift the median redshift of star formation to higher $z$. This
will tend to decrease the in-situ fraction. Thus, improved feedback
should bring our results and those of Os10 into better agreement. }

In addition to differences in resolution {\rm and feedback}, differences in star
formation efficiency between the two sets of simulations affect the
fraction in-situ versus accreted stars. \citet{Naab07} have
shown that the {\it total} stellar mass is not sensitive to the star
formation timescale. However, in simulations with short star formation
timescales, stars will form in small clumps outside the galaxy
and will be later accreted, while in galaxies with longer star
formation times, small clumps will instead be accreted as gas and later
form stars in-situ. Thus, the total number of baryons accreted
by the galaxy remains the same, but whether the baryons are accreted as stars
or gas depends strongly on the star formation timescale. {\rm This
  effect is at least comparable to the effects from feedback on the
  accreted fraction. In comparing two grid-based codes with different
  star formation efficiencies, the redshift by which half the stars
  have formed moves from $z\sim2$ for an efficiency of 5\% to
  $z\sim1$ for an efficiency of $1\%$ \citep{Scannapieco2011},
  comparable to the change in the median redshift of star formation
  from different feedback models.} 

Ignoring the
return of mass to the gas phase from high-mass stars, the star
formation rate can be written as 
\begin{equation}
\label{eq:rycSFR}
\frac{d M_*}{d t} = C_* \frac{M_{\mathrm{gas}}}{t_{\mathrm{dyn}}}\  ,
\end{equation}
where $t_{\mathrm{dyn}}$ is the dynamical time of the gas forming
stars (see equation \ref{eq:sfr_crit}). Thus, $1/C_* \times
t_{\mathrm{dyn}}$ is the star formation timescale,
$t_{\mathrm{SF}}$. Simulations with
large values for $C_*$ have short $t_{\mathrm{SF}}$ and therefore make
stars efficiently. In the simulations examined in this work, $C_* =
0.03$, in agreement with theoretical calculations \citep{Krumholz05}
and observations \citep{Kennicutt98,Evans09}. In Os10, the value for
$C_*$ is $0.083$, which means for a given density gas clump, star
formation is $\sim 2.8$ times faster in the simulations used by
Os10. This difference in star
formation rate becomes especially important for small clumps of
baryons being accreted onto larger galaxies. 

We can compare the star
formation time scale to the timescale for material to fall in from the
virial radius. The infall time is given by the dynamical time of the
dark matter halo, $t_{\mathrm{virial}} \propto
1/\sqrt{G\rho_{\mathrm{virial}}} = 1/\sqrt{200
  G\rho_{\mathrm{crit}}}$. Thus, the ratio of the star
formation time to the infall time is given by
\begin{eqnarray}
\label{eq:TsfrTinfall}
t_{\mathrm{SF}} / t_{\mathrm{virial}} &=& \frac{\sqrt{200
    G\rho_{\mathrm{crit}}}}{C_*\sqrt{G\rho_{\mathrm{gas}}}} \\
\nonumber
&=& \frac{1}{C_*}\sqrt{\frac{300}{4\pi}}H_0\ t_{\mathrm{dyn}}\\ \nonumber
&=& 0.05h\left(\frac{1}{C_*}\right)\left(\frac{t_{\mathrm{dyn}}}{100\
    \mathrm{Myr}}\right)\ .
\end{eqnarray}
For $t_{\mathrm{dyn}} = 100\ \mathrm{Myr}$, the ratio of the star
formation timescale to the infall timescale is $0.6\ h$ and $1.7\ h$ for $C_* =
0.083$ and $0.03$, respectively. In the first case, an infalling clump
of gas above the star formation threshold density 
will convert 
$\sim 91\%$ of its gas to stars, while in the second case only $57\%$
of the gas will be converted into stars, and the accreted fraction
will decrease by about $40\%$. The difference in $C_*$ between the two
simulations partially explains the difference of a factor of $2$
between the accreted fractions found in this work and Os10. It also
highlights the importance of the star formation efficiency in studying
the merger histories of galaxies. Unlike dark matter halos, which always
undergo `dry' mergers, baryons can merge either as stars or gas. The
relative fraction of stars to gas can have profound effects on the
observable properties of galaxies.

\subsection{Properties of accreted and in-situ stars}
\label{ssec:props_AccIs}
\figAgeMassGrp
\figZMassGrp

In addition to examining the total contribution of in-situ and
accreted stars, we can also examine the mean properties of the
accreted and in-situ fractions as a function of final galaxy stellar
mass. The top panels in Figures \ref{fig:m_agesISACC} and
\ref{fig:m_zsolISACC} show the mass-weighted stellar ages and stellar
metallicities for the in-situ and accreted stars as a function of
$z=0$ galaxy stellar mass. It is immediately evident that the accreted
stars are older (by $2-2.75$ Gyrs) and more metal-poor (by $\sim 0.15$ dex)
than the stars 
formed in-situ. Figure \ref{fig:mGroup_ages} shows distributions of
stellar ages for galaxies divided into three stellar mass
bins. The 
vertical lines show the median ages for the in-situ, accreted, and
total stellar masses. Again, it is clearly
evident from these figures that the accreted stars are older than the
in-situ stars. The plots in this figure as well as Figures
\ref{fig:mGroup_metals}$-$\ref{fig:envGroup_age} weight all galaxies
in the cluster and void boxes equally. If we weight the galaxies
according to the fit to the star formation history \citep{Cen2011},
only the results for the lowest mass bin change, and the differences
in the median values are very small. Furthermore, by dividing the galaxies into mass bins,
we see in Figure \ref{fig:mGroup_ages} that there is more late-time star formation in lower mass ($M_* \la
10^{11}\ \Msun$) than in higher mass galaxies, consistent with the
results of \citet{Cen2011} using the same simulation presented
here. {\rm Examining the trends in median stellar age for the accreted
  and in-situ fractions as a function of mass reproduces the trends
  shown in Figure \ref{fig:m_agesISACC}; namely, while the in-situ and
  accreted stellar ages only change slightly with stellar mass, the
  median age of all the stars increases since the accreted fraction
  grows with stellar mass. Finally, the difference in
  stellar age between the accreted and in-situ stars is roughly
  independent of total stellar mass. If we exclude galaxies more
  massive than $10^{11.1}\ \Msun$ and only examine the top panel, the
  accreted stars are still more than $2$ Gyrs older on average than
  stars formed in-situ. }

Since the age
distributions of the in-situ 
and accreted stars are different, it is not surprising that the
metallicity distributions are different as well (see Figure
\ref{fig:mGroup_metals}). For the most massive galaxies, the accreted
stars are $0.12$ dex more metal-poor than the stars formed
in-situ. {\rm The sign of this difference is in agreement with simulation
results from \citet{Font2011} for disk galaxy halos, but the
metallicity difference found in this work is smaller.} However, as noted above, the stellar metallicities in the
simulation are sensitive to the supernovae feedback prescription, and are highly
uncertain. Nonetheless, it is unlikely that improved feedback
would reverse the sign of the difference in metallicity between
accreted and in-situ stars. {\rm Additionally, the difference in
  metallicity between accreted and in-situ stars is largest for the
  lowest stellar mass bin, where the mass function of galaxies agrees
  with observations. However, additional supernovae feedback may play
  an important role here, as well. }

The bimodal
distribution of the stellar metallicities in the low mass galaxies
(see the top panel of Figure
\ref{fig:mGroup_metals}) may be partially due to over-weighting the cluster
galaxies relative to the void galaxies. If we weight the galaxies
as in \citet{Cen2011} (the void galaxy contribution is increased by a
factor of $6$), the metallicity distribution in the top panel of
Figure \ref{fig:mGroup_metals} is still bimodal, but the low
metallicity peak is nearly equal in height to the high metallicity
peak. This remaining bimodality may be due to enhanced late star
formation in metal-rich gas, which is not observed in present-day
galaxies, and probably due to inadequate resolution and stellar
feedback. 

{\rm In addition to differences in the median stellar
  metallicities for accreted and in-situ stars, the dispersions are
  also different; the accreted stars have a larger metallicity
  dispersion. For the three panels in Figure \ref{fig:mGroup_metals},
  the inter-quartile ranges of the in-situ stellar $Z/Z_\odot$ are
  $0.26$, $0.28$, and $0.27$ dex. The same ranges for the accreted stars
  are $0.60$, $0.41$, and $0.34$ dex. The differences in metallicity
  distribution width are due to the
  inhomogeneous make-up of the accreted material. The accreted stars
  are formed in a variety of systems with different metallicities,
  while the in-situ stars are formed in a single stellar system. The same
  difference in metallicity dispersion is seen in the Galactic disk
  and halo 
  \citep[e.g.][]{Beers1985} and is used as evidence for the accretion
  origin of the Galactic halo \citep[e.g.][]{SearleZinn78, Unavane96}.}

We also examine the spatial distribution of in-situ and accreted stars
in galaxies at $z=0$. Figure \ref{fig:mGroup_profile} shows the
$3$-dimensional average stellar mass profiles for the accreted and
in-situ stars in 
three different mass bins. As expected (and as shown in the work of \citet{Oser11}), the half-mass radius of the
accreted stars is larger than the half-mass radius of the in-situ
stars by approximately a factor of $2$ for all three mass bins. {\rm
 Some of the accreted material on the outskirts of galaxies may be due
 to our choice of density threshold in the HOP grouping algorithm; a
 lower threshold will lead to larger galaxies with larger accreted
 fractions. In order the check how much this affects our results, we
 show the median radius of the stars accreted from {\it resolved}
 galaxies only in Figure \ref{fig:mGroup_profile}. This radius is
 still larger than the median radius of the in-situ stars. Furthermore,
 this radius represents the minimum half-mass radius for the accreted
 material, as additional accretion from unresolved galaxies will tend
to increase this radius.}

Trends with accreted fraction and radius suggest that the average
merger event is not an equal mass-merger, 
after which the stars would be well-mixed, but rather 
a minor merger in which low density material is added to the
outside of a massive galaxy. It also suggests that minor mergers can
substantially increase the half-light radius of massive galaxies,
leading to the growth of early-type galaxies at late times. This is 
in agreement
with both observations \citep{Bezanson09,vanDokkum2010, Newman2012,
  Tal2012} and other simulations \citep{Gallagher72,
  BoylanKolchinMa04, Naab07, Naab09, Oser11}. {\rm It also follows
  from simple,
semi-quantitative arguments based on the virial theorem
\citep{Naab09, Bezanson09}.} Because the accreted
material has lower metallicity 
than the in-situ material, we expect
galaxies to have negative metallicity and colour gradients. In
projection, the average metallicity gradient for the galaxies with
robust merger histories is $\dif \log (Z/Z_\odot)/\dif\log(R/\Reff) =
-0.38\pm 0.23$, where \Reff{} is the projected half-mass radius. {\rm
  Furthermore, the differences in the distributions of stellar
  metallicity between the accreted and in-situ stars lead to an
  increase in the metallicity dispersion as a function of radius. The
  standard deviation of $\log Z/Z_\odot$ has an average slope of $\dif 
\sigma_{\log Z/Z_\odot}/\dif\log(R/\Reff) = 0.28 \pm 0.18$. There are
many mechanisms besides accretion that will contribute to an increase
in metallicity dispersion. Indeed, we find the in-situ stars
alone also have an increase in metallicity dispersion with radius only
slightly smaller than the increase reported above. }

Observations of these
gradients are difficult because the low surface brightness in the
outskirts of galaxies. There are numerous measurements of
colour gradients in early-type galaxies \citep{deVaucouleurs61,
  Eisenhardt2007, Tortora2010, Suh2010, Tal2011, LaBarbera2011, Guo2011,
  GonzalezPerez2011}, although only a few studies extend beyond the
half-light radius of 
the galaxy. Furthermore, colour measurements cannot be used to
distinguish between metallicity and age gradients. More recently,
studies of the metallicity gradients of globular cluster systems in 
early-type galaxies support theories of two-phase galaxy formation \citep{Arnold2011,Forbes2011}. Direct measurements of metallicity
gradients in the stellar populations of early type galaxies have been
done using long slit spectroscopy \citep{Davies1993, Mehlert2003,
  Brough2007, Spolaor2010}, and integral-field-units \citep{Rawle2008,Kuntschner2010, Greene2012}. Only a few 
measurements focus on gradients beyond the half-light radius
\citep{Faber1977, Foster2009, Spolaor2010, Greene2012}. These studies
find metallicity gradients (but \emph{not} age gradients) which
indicate that accretion of low metallicity systems is   
important in the build-up of massive galaxies \citep{Foster2009, Spolaor2010, Greene2012}. \citet{Greene2012} find an
average slope of $\mathrm{[Fe/H]} \approx -0.1$ from $0\sim2.5$ \Reff{}
for $8$ massive early type galaxies, in good agreement with the
average slope we measure in the simulations. Using the iron and
magnesium abundances in the outskirts of galaxies, 
they obtain an average mass ratio for the merger events of $10:1$,
which agrees roughly with our estimate for the mean merger ratio below
(see \S\ref{ssec:mmratios}).
\figProfileMass

Although the accreted material is made up of older stars, it is not
added to the galaxies until late times. Figure
\ref{fig:mGroup_assemble} shows the mass-weighted average mass
assembly for galaxies in three mass bins. For the central mass bin,
half of the accreted stars are formed before $z\approx 1.6$, but it
takes until $z \approx 0.6$ for half of these stars to actually be
accreted. Furthermore, Figure \ref{fig:mGroup_assemble} shows that the
in-situ stars are in place before the accreted material is
added.  The difference in assembly times between the accreted and
in-situ stars is largest for the lower mass galaxies. This is
consistent with the 
picture of two phase galaxy formation \citep{Oser10, Naab07}; 
galaxies first undergo a phase of intense, in-situ star formation,
followed by a phase of accretion of old, less massive, and therefore more
metal-poor systems.

\figBuildMassGrp

\subsection{Mergers and environment}
\label{ssec:mergerenviron}
We can also use the galaxies from the cosmological simulations to examine the
properties of mergers as a function of environment. This analysis is not
possible for simulations which only include individual
halos. Studies of paired galaxies have found a dependence on the
merger rate with environment \citep{McIntosh2008, Darg2010, Lin2010,
  Tonnesen2011}. This is expected, as higher density environments have
more galaxies and higher potential
for merging. However, part of this trend may be due to the correlation
of environment and galaxy mass, which is related to the accreted
fraction. By selecting a sample of galaxies of similar mass, we can
eliminate trends in the accreted fraction with mass and 
isolate the 
trends with environment.  

We define environment as the
3-dimensional galaxy density to the $5^{\mathrm{th}}$ nearest neighbor
($\rho_5 = 6/(\mathrm{distance\ to\ 5^{th}\ neighbor})^3\ [\h^3\Mpc^{-3}]$). The
densities used are an average of the $z=0$ and $z=0.05$ measured
densities in order to eliminate some of the noise introduced by the
nearest neighbor density measure. We have restricted the sample
to $109$ galaxies with stellar masses between $6\times 10^{10}\ \Msun < M_*
< 3 \times 10^{11}\ \Msun$. Overall, we find that the accreted
fraction is not a 
strong function of present-day environment; the average accreted
fraction increases from 
$\sim 0.20$ for field galaxies to $\sim 0.25$ for cluster galaxies. This 
contradicts the observations of
\citet{McIntosh2008}, which show an increase in the merger rate for
group galaxies by a factor of 
2--9 over the rate for field galaxies. 
The difference is due in part to
our decision to study trends with environment at constant stellar
mass, eliminating the larger trend. 
Furthermore, \citet{McIntosh2008}
use group and cluster membership as a measure of environment. This
measure is more sensitive to the near-field environment of a 
galaxy than $\rho_5$, and therefore, it is probably more closely
correlated with the number of mergers a galaxy has undergone.

Figure \ref{fig:envGroup_age} shows the stellar
age distributions for galaxies in three density bins, which correspond
to the local density for field, group, and cluster
galaxies. {\rm Typical separations between galaxies in each environment
  bin are $5\ \Mpch$, $1\ \Mpch$, and $0.5\ \Mpch$, respectively.} These bins were chosen to have equal numbers of galaxies in the
mass range $6\times 10^{10} \Msun < M_*
\leq 3 \times 10^{11} \Msun$. However, the results are not sensitive to
the exact locations of the bin separations. As expected, the star formation
histories of the galaxies in low density environments are far more
extended. For field galaxies, both the in-situ and accreted stars are
younger than in group and cluster galaxies by $\sim2$ Gyr. In fact,
not only does the in-situ star formation occur later in low density
regions, but the accretion also occurs $\sim 1.5$ Gyr later than in
the higher density regions. In every respect, low density regions
evolve more slowly than high density regions. 

Figure 
\ref{fig:envNormed_age} shows the total star 
formation histories for galaxies in low, medium, and high density
regions. These
histories are normalized to the local dark matter mass in each region,
which is accomplished by weighting each galaxy by the dark matter mass
enclosed in a sphere centered on the galaxy and extending to the
galaxy's fifth nearest neighbor. As expected, star formation in galaxies in the
lowest density environments dominates today, while, at high redshift,
star formation occurred mainly in the mid- and high density
regions. This is consistent with the star formation histories
presented  in \cite{Cen2011}. 


\figEnvAge
\figEnvNormAge

\section{Accreted stars and Mergers}
\label{sec:mergers}
The accreted fraction of the final stellar mass is added to the
galaxies through mergers. Using the simulations we can determine the
rate at which mergers occur and the 
distribution of mass ratios 
of these
mergers. Since the accreted fraction in most of our galaxies is less
than $30\%$ we expect that minor mergers (merger mass ratios less than
$4$:$1$) will dominate the accretion history of galaxies in our
sample. In the following sections, we compute the merger rate for our
sample as a function of redshift and the average merger ratios for the
most massive galaxies in the simulation.

\subsection{Merger rates}
\label{ssec:merger_rates}
The slopes of the in-situ and accreted mass
assembly histories in Figure \ref{fig:mGroup_assemble} are 
the star
formation rates and merger rates, respectively. The
star formation rate for the massive galaxies 
($\log M_*/\Msun > 12.2$) peaks
at $z\approx2.5$, when the star formation rate is high enough to
double the galaxy mass within a Gyr. The accretion rate for the
massive galaxies peaks later, at $z\approx 1$, with an average growth
rate from accretion of $\sim 15\%\ \mathrm{Gyr}^{-1}$. At $z=0.25$, the
galaxies with $M_* \ga 3\times 10^{11}\Msun$ are adding stellar mass
in mergers at a rate of $\sim 2.6\%\ \mathrm{Gyr^{-1}}$. This includes
growth from unresolved mergers in the simulations. 

The simulation merger rate is approximately a factor of two
larger than the observed merger rate for luminous red galaxies (LRGs) of 
$1.7\% h\
\mathrm{Gyr}^{-1}$, where $H_0=100h\ \kmps\Mpc^{-1}$, found by \citet{Masjedi08} using the small scale
cross correlation function to determine the close pair fraction in
SDSS. This observed rate includes growth from minor mergers since the maximum
luminosity ratio between the LRGs and other galaxies is $\sim 50$:$1$
in SDSS observations \citep{Masjedi08}. Since the massive galaxies in
the simulation suffer from overmerging, we expect the simulation
merger rate to be inflated compared to the real value. 

Comparing our simulated merger rates to other observed merger rates
based on paired galaxies or morphologically-disturbed galaxies is not
straightforward; it requires knowing the average merger mass ratio and the
merger rate of a single population of galaxies as a function of
time. Assuming an average merger mass ratio between $0.15$ and $0.3$
(compatible with our results from \S\ref{ssec:mmratios}), the
merger rate as measured by quantified morphological disturbances ($G-M_{20}$) in
\citet{Lotz2011} is $7.5-15\%\ \mathrm{Gyr^{-1}}$ at $z=0.2$ for galaxies at
number densities above $\sim 6\times 10^{-3}\
\mathrm{Mpc^{-3}}$. This
is considerably larger than both the merger rate measured in our
simulations and that measured by \citet{Masjedi08}, suggesting that
the observational merger rate is still uncertain.  Using
higher resolution simulations with the same initial conditions,
\citet{Cen2011b} find that the observationally-defined merger rate
from close pairs of galaxies does exceed the physical merger rate for
simulated galaxies. This over-estimate is at least partially corrected
for by including an observable timescale for the merger
\citep{Lotz2011}. Despite this correction, the observed merger rates
are still large. The \emph{total} merger
rate from our simulations is 
comparable 
to the observed \emph{major} merger rates from studies of close
galaxy pairs \citep[e.g.][]{Bundy2004,Kartaltepe2007,deRavel2009,Bundy2009}; visible disturbances (tidal tail, bridges,
etc.) \citep{Jogee2009,Darg2010,Bridge2010, Kartaltepe2010}, and
quantified morphological disturbances \citep[e.g.][]{Lotz2008,
  Conselice2009,   LopezSanjuan2009}, respectively.


We can also compare the evolution of the merger rate as a function of
redshift to the observations of merger rate evolution. However, care must be taken to
ensure the selection of galaxies from the simulations and observations
are similar. \citet{Lotz2011} argue that galaxies selected based on
luminosity or mass cuts do not give the same evolution as samples
selected based on constant galaxy number density, as the progenitor
and descendant galaxies are different in the mass and luminosity selected
samples. In our simulations, the merger rate per galaxy per Gyr 
for galaxies with stellar masses above $10^{10}\ \Msun$ grows as
$(1+z)^{2.0\pm0.25}$ for $z<1.5$. Thus, we expect the merger rate to
be an increasing function of redshift, in agreement with observations
\citep[but see \citealp{Bundy2009}]{Bridge2010}.  The slope of
the galaxy merger rate with redshift is similar to the slope of the dark
matter merger rate with redshift \citep[e.g.][]{Gottlober2001,
  Fakhouri2008, Fakhouri2010}. However, the simulations clearly show
systems in which the  dark
matter halos 
merge well before the galaxies merge. Cosmological hydrodynamical
simulations, such as those used here, can be used to measure
the time lag between dark matter halo mergers and galaxies mergers,
which we plan to pursue in future work.

\subsection{Expected merger ratio}
\label{ssec:expectedMR}
In addition to studying the rate of merging, the simulations also
allow us to examine the average merger mass ratios as a function of galaxy
mass. We define the merger ratio, $\mu$, as the mass of the accreted
galaxy divided by the mass of the progenitor galaxy. By definition,
$\mu$ is always less than one. We first calculate the 
expected merger ratio given the galaxy mass function, and compare this
expected value with the value from simulations. Similar calculations
for the limiting case of small merger ratios are done in Os10 and
\citet{Naab09}. 

Ignoring effects of
dynamical friction, gravitational focusing, and local density, the
mean merger mass ratio for a galaxy is the average mass of less massive
galaxies divided by the parent galaxy mass. This can be calculated from the mass function. Given a mass
function $\Phi(M)\dif M$, the probability of a galaxy with mass $M_0$
accreting a galaxy with mass in the range $(M_a,M_a+\dif M)$ is
\begin{equation}
\label{eq:probAccrete}
  P(M_a)\dif M = \frac{\Phi(M_a)\dif
    M}{\int_{M_{\mathrm{min}}}^{M_0}\Phi(M)\dif M}\ ,
\end{equation}
and the number-weighted and mass-weighted average merger mass ratios are
simply
\begin{eqnarray}
\label{eq:ratioAccrete}
\langle\mu_n\rangle &=& \int^{M_0}_{M_{\mathrm{min}}}
m P(m)\dif m /\int^{M_0}_{M_{\mathrm{min}}} P(m) \dif m \nonumber\ \mathrm{,\ and} \\ 
\langle\mu_m\rangle &=& \int^{M_0}_{M_{\mathrm{min}}} m^2P(m)\dif m / \int^{M_0}_{M_{\mathrm{min}}} m P(m)
\dif m\ ,
\end{eqnarray}
respectively. For a Schechter function, $P(M) \propto
(M/M_*)^\alpha\exp^{-M/M_*}$, and the number-weighted and
mass-weighted average merger ratios can be
evaluated as
\begin{eqnarray}
\label{eq:ratioAccreteSchecht}
\mun &=& 1/M_0 \times
\left(\Gamma\left(\alpha+2,M_{\mathrm{min}}\right)
  -\Gamma\left(\alpha+2,M_0\right)\right)/\nonumber \\
&&\left(\Gamma\left(\alpha+1,M_{\mathrm{min}}\right)-
  \Gamma\left(\alpha+1,M_0\right)\right) \nonumber\ , \\
\mum &=& 1/M_0 \times \left(\Gamma\left(\alpha+3,M_{\mathrm{min}}\right) -
  \Gamma\left(\alpha+3,M_0\right)\right)/\nonumber \\
&&\left(\Gamma\left(\alpha+2,M_{\mathrm{min}}\right)- \Gamma\left(\alpha+2,M_0\right)\right)\ ,
\end{eqnarray}
where masses are given in units of $M_*$, and we have included a lower
limit on the mass function, $M_{\mathrm{min}}$. The dashed
lines in Figure \ref{fig:m_mergerratio} show the numerical values for
\mun and \mum for
$\alpha=-1.16$ and $\log M_*/\Msun = 10.8$, the Schechter function
parameters fit to the SDSS sample \citep{Li2009}. The integrals are
stopped at and $M_{\mathrm{min}}=10^9\ \Msun$. This procedure
yields similar results to the expected merger mass ratio calculated by
\citet{Hopkins10} using N-body simulations for the merger trees
and semi-empirical methods for determining the galaxy
properties. \citet{Hopkins10} find that \mum{} peaks around
$M_*\approx 10^{11}\ \Msun$, and decreases for both larger and smaller
galaxy masses. 

\figmergerratio

\subsection{Mean merger mass ratio}
\label{ssec:mmratios}
In \S\ref{ssec:galaxyPop} we show that the simulated mass function
deviates from the observed mass function below $M_* <
10^{10}\Msun$. Therefore, in this section, we only examine mergers
between galaxies with stellar masses greater than $10^{10}\ \Msun$ at
$z < 2.5$. We require that the final galaxy mass is larger than
$10^{11}\ \Msun$, thus allowing us to resolve mergers with a mass
ratio of $\mu < 0.1$ at $z=0$. {\rm This limits the sample to
  the most massive galaxies which suffer from over-cooling and too much
  star formation, making comparisons to observations ineffective.} This sample includes $85$ galaxies
from the cluster box and $15$ galaxies from the void box. Figure
\ref{fig:m_mergerratio} shows the mean mass-weighted and
number-weighted merger ratios as a function of $z=0$ stellar mass. The
large triangles denote median merger ratios and do not show a
significant trend 
with galaxy mass, although the maximum merger ratio increases with
decreasing stellar mass. The averages are consistent with the expected
merger ratio from the mass function in the simulation (solid
line). The excess of massive galaxies {\rm due to over-cooling} makes 
the expected merger ratios in the simulation deviate 
from the merger ratios predicted by a Schechter function galaxy mass
distribution (cf. Figure \ref{fig:massFunc}). The 3-$\sigma$ clipped average values for \mun{} and \mum{}
are $0.20\pm 0.16$ and $0.26\pm 0.21$, respectively. These results are 
roughly consistent with the merger ratios calculated in
\citet{Hopkins10} using semi-empirical techniques.

\citet{Oser11} examine the average merger ratio for the same set
of galaxies used in Os10. They also find little trend with merger
ratios and stellar mass, but their average merger ratios are lower
than the ones we find here and closer to the analytically derived values. Because the simulations in Os10 are of higher
mass resolution, they are able to resolve mergers down to 
lower
mass ratios. Therefore, it is not surprising that they find a substantially
lower value for \mun{}, which is affected by many low mass mergers,
while their value for \mum{} is comparable to the \mum{} we
measure. Because we are restricted to mergers with mass ratios greater
than $10$:$1$, the plots in Figure \ref{fig:m_mergerratio} only account
for $75\%$ of the accretion onto the galaxies shown. The other $25\%$
of the accretion is from galaxies below our mass cutoff and other unresolved
accretion. Accounting for this accretion sets a lower limit on \mum{} of
$0.20$, in agreement with the results from \citet{Oser11}. Finally, the
low value for \mun{} reported by \citet{Oser11} is consistent with
the differences in star formation efficiency in the two works; the
higher star formation efficiency in Os10 and \citet{Oser11} should
yield more small stellar systems than in our simulations, which will
decrease \mun{}.

%
\section{Discussion and Summary}
\label{sec:summary}
\figmhostz

The two-phase picture of galaxy formation put forward by Os10 provides
a useful framework for 
studies of galaxy 
evolution. {\rm We apply 
this framework to study the galaxy
merger histories of more than $600$ simulated galaxies with $M_* >
10^{10}\ \Msun$. In this work, we corroborate 
the two-phase model for galaxy formation.} The first phase consists of
a period of in-situ star formation, which occurs around $0.75<z<1$. This
in-situ phase accounts for 
60--90\% 
of the star formation in
galaxies with stellar masses above $10^{10}\ \Msun$. The remainder of
the stars are added in mergers. The peak of the merger activity occurs
at $z\approx0.5$, after the majority of the in-situ star formation has
taken place. {\rm Our results are in good agreement with those from
  Os10 and other simulations distinguishing between accreted and
  in-situ star formation in stellar halos, mainly around disk galaxies
  \citep[e.g.][]{Font2011, Abadi2006, Zolotov2009,
    Zolotov2010}. In particular, the differences in stellar age,
  metallicity, and spatial distribution shown in Figures
  \ref{fig:mGroup_ages}--\ref{fig:mGroup_profile} are in agreement
  with the differences found by \citet{Font2011} and \citet{Zolotov2009}.}

Although the accretion occurs at late times, the
accreted systems are old. The accreted stars are on average
$2$ Gyr older than the stars formed in-situ. This age difference
correlates with a median  metallicity difference of $0.15$ dex. {\rm Since the
accreted material comes from a variety of sources (i.e., a range of
galaxy host masses), the distribution of stellar metallicities is
larger for the accreted stars than for the in-situ stars.} However, the
simulated galaxies do not conform to the mass-metallicity relation, 
suggesting 
that the stellar feedback model does not provide sufficient
feedback, and 
that higher resolution is needed. Improved feedback 
prescriptions 
should
lower the stellar metallicities for 
the less massive galaxies and the accreted systems, thus increasing
the differences in metallicity between in-situ and accreted
stars. This effect may be countered by a decrease in metallicity
in the centers of galaxies, if higher resolution and better feedback 
prescription help shut off the excess late star formation currently seen in the
simulations. 

Unlike 
in 
simulations, observations cannot easily distinguish between
stars formed in-situ and those added by accretion. Nonetheless, we find
indirect signs of the two phases of galaxy formation which are
observable today. Because the accreted stars formed in small systems
at early times, their stellar ages and metallicities are noticeably
different from the stars formed in-situ. Furthermore, because the
accreted stars preferentially reside 
in the outskirts of galaxies, we
find strong gradients in the stellar populations of massive galaxies
which have been observed beyond the half light radius of early-type
galaxies \citep{Foster2009,Spolaor2010, Greene2012}.

As explained above, our results 
are consistent with 
models 
of 
hierarchical
formation of galaxies; large systems are assembled at late times from
older, smaller building blocks. Figure \ref{fig:mhostz} is an
illustration of 
the 
hierarchical growth. Here, we plot the stellar age
against the stellar mass of the system (galaxy) 
in which each star was born, 
for the accreted and in-situ stars, separately. For a galaxy with no major
mergers (central panel), the distribution of accreted and in-situ
stars are well separated. On average, accreted stars form early in
small systems, while in-situ star formation occurs later, when the
host galaxy is more massive. On the other hand, the distributions of
accreted and in-situ stars are very similar for a galaxy which
underwent a major merger (left panel). This is unsurprising since two
galaxies with roughly the same mass should have similar star formation
histories.

{\rm The overall trends in Figure \ref{fig:mhostz} suggest that
galaxy stellar mass and stellar age are anti-correlated. However, the low mass
galaxies shown in this figure formed early and were subsequently accreted, at
which point they stopped forming stars. These galaxies are not
representative of present-day low mass galaxies, but are
rather the building blocks of more massive galaxies. Indeed, the top
panel of Figure \ref{fig:m_agesISACC} shows stellar mass and stellar
age are correlated for galaxies which survive to
 $z=0$, in qualitative
agreement with observations. }

Observations cannot directly reproduce Figure
\ref{fig:mhostz}. However, 
there may be a close mapping from the relation between stellar ages and progenitor 
masses to the relation between metallicity and $\alpha$-element 
enhancement \citep{Thomas05, Johnston08}. Old stellar populations have higher
alpha-to-iron abundance ratios than younger stellar systems, while massive
galaxies typically have higher metallicities than less massive
systems. {\rm Figure \ref{fig:mhostz} shows that accreted stars come
  from older, less massive systems, so we expect that accreted stars
  should be alpha-element enhanced compared to stars formed
  in-situ. This will yield radial gradients in abundance ratios in addition
  to the gradients in metallicity discussed above. } This stellar abundance space 
is used to map the accretion
history of the Galactic halo
\citep{BlandHawthorn03,Robertson05,Font06a,Font06b,Johnston08,Zolotov2010}. Using 
simulations of a Milky Way analogue, \citet{Tissera2012} find that the
accreted stars in the disc ($15\%$ of the disc stars) are alpha-enhanced
and older than the stars formed in-situ, in 
broad agreement with our
results shown in Figure \ref{fig:mhostz}. This type of analysis is
being extended to galaxies besides the Milky Way by measuring 
$\alpha$-to-Fe
ratios as a function of galaxy radius \citep{Kuntschner2010, Spolaor2010, Greene2012}. As detailed chemical and spatial information
becomes available for the outer regions of more  galaxies, it may be
possible to differentiate between a minor merger and major merger
driven galaxy formation history for individual galaxies.  

The results of this work agree broadly with the results of Os10 based
on a set of $40$ galaxies simulated using GADGET, an SPH code, at
higher mass resolution. We claim that most of the differences can be
accounted for by the differences in star formation efficiency, feedback, and
resolution between the simulations. 
The higher 
star formation efficiency in
Os10 yields a higher accreted fraction because small halos have time to 
form stars during infall from the virial radius of larger halos. These
small stellar systems tend to decrease the
number-weighted mean merger ratio \citep{Oser11}. In our simulations, the lower
resolution tends to increase the in-situ star
formation rate  by preventing gas clumps from
fragmenting and forming stars before being accreted. These gas clumps
survive 
to 
the centers of the galaxies, where they then form stars
in-situ. The effects of resolution on the in-situ and accreted
fractions will be checked in future AMR simulations run at higher
resolution. Together with the differences in feedback, these effects account for the difference of a
factor of two between the accreted fraction observed in the SPH and
AMR simulations. Despite these differences, there is an excellent qualitative
agreement between this work and the work in Os10. This shows 
that 
the two phase model for the formation of massive galaxies is robust and
independent of the numerical methods and subgrid physics. {\rm
  Therefore, the simulations can be used to predict observable consequences
  of the two phases of galaxy formation. In this work, we show that
  the differences in age, metallicity, and metallicity dispersion
  between the accreted 
  and in-situ stars yield radial gradients in these quantities for
  present-day massive galaxies. Currently, there is evidence for
  radial trends in metallicity in early-type galaxies \citep[cf.][]{Greene2012,
    Spolaor2010}, consistent with our predictions for two phase galaxy
formation and further observational tests of this picture will be
 extremely useful. }

\section*{Acknowledgments}
We are very grateful to the referee whose insightful comments greatly
improved this manuscript. We 
also thank Greg Bryan for help with Enzo code, and Ludwig Oser for
useful comments and providing data for comparison. Computing
resources were in part provided by the NASA High--End Computing (HEC)
Program through the NASA Advanced 
Supercomputing (NAS) Division at Ames Research Center.
This work is supported in part by grants NNX11AI23G. CNL acknowledges
support from the NSF grant AST0908368.


\bibliographystyle{mn2e}
\bibliography{lackner_short}

\begin{thebibliography}{}

\bibitem[\protect\citeauthoryear{{Abadi}, {Navarro} \& {Steinmetz}}{{Abadi}
  et~al.}{2006}]{Abadi2006}
{Abadi} M.~G.,  {Navarro} J.~F.,  {Steinmetz} M.,  2006, \mnras, 365, 747

\bibitem[\protect\citeauthoryear{{Agertz}, {Moore}, {Stadel}, {Potter},
  {Miniati}, {Read}, {Mayer}, {Gawryszczak}, {Kravtsov}, {Nordlund}, {Pearce},
  {Quilis}, {Rudd}, {Springel}, {Stone}, {Tasker}, {Teyssier}, {Wadsley} \&
  {Walder}}{{Agertz} et~al.}{2007}]{Agertz07}
{Agertz} O., et al. 2007, \mnras, 380, 963

\bibitem[\protect\citeauthoryear{{Arnold}, {Romanowsky}, {Brodie}, {Chomiuk},
  {Spitler}, {Strader}, {Benson} \& {Forbes}}{{Arnold}
  et~al.}{2011}]{Arnold2011}
{Arnold} J.~A., et al. 2011, \apjl, 736,   L26+

\bibitem[\protect\citeauthoryear{{Beers}, {Preston} \& {Shectman}}{{Beers}
  et~al.}{1985}]{Beers1985}
{Beers} T.~C.,  {Preston} G.~W.,  {Shectman} S.~A.,  1985, \aj, 90, 2089

\bibitem[\protect\citeauthoryear{{Behroozi}, {Conroy} \& {Wechsler}}{{Behroozi}
  et~al.}{2010}]{Behroozi2010}
{Behroozi} P.~S.,  {Conroy} C.,  {Wechsler} R.~H.,  2010, \apj, 717, 379

\bibitem[\protect\citeauthoryear{{Bezanson}, {van Dokkum}, {Tal}, {Marchesini},
  {Kriek}, {Franx} \& {Coppi}}{{Bezanson} et~al.}{2009}]{Bezanson09}
{Bezanson} R.,  {van Dokkum} P.~G.,  {Tal} T.,  {Marchesini} D.,  {Kriek} M.,
  {Franx} M.,    {Coppi} P.,  2009, \apj, 697, 1290

\bibitem[\protect\citeauthoryear{{Bland-Hawthorn} \&
  {Freeman}}{{Bland-Hawthorn} \& {Freeman}}{2003}]{BlandHawthorn03}
{Bland-Hawthorn} J.,  {Freeman} K.~C.,  2003, in {E.~Perez, R.~M.~Gonzalez
  Delgado, \& G.~Tenorio-Tagle} ed., Star Formation Through Time Vol.~297 of
  Astronomical Society of the Pacific Conference Series, {Unravelling the Epoch
  of Dissipation}.
p.~457

\bibitem[\protect\citeauthoryear{{Blanton}, {Hogg}, {Bahcall}, {Brinkmann},
  {Britton}, {Connolly}, {Csabai}, {Fukugita}, {Loveday}, {Meiksin}, {Munn},
  {Nichol} et~al.,}{{Blanton} et~al.}{2003}]{Blanton03b}
{Blanton} M.~R., et al. 2003, \apj, 592, 819

\bibitem[\protect\citeauthoryear{{Boylan-Kolchin} \& {Ma}}{{Boylan-Kolchin} \&
  {Ma}}{2004}]{BoylanKolchinMa04}
{Boylan-Kolchin} M.,  {Ma} C.,  2004, \mnras, 349, 1117

\bibitem[\protect\citeauthoryear{{Bridge}, {Carlberg} \& {Sullivan}}{{Bridge}
  et~al.}{2010}]{Bridge2010}
{Bridge} C.~R.,  {Carlberg} R.~G.,  {Sullivan} M.,  2010, \apj, 709, 1067

\bibitem[\protect\citeauthoryear{{Brook}, {Kawata}, {Gibson} \&
  {Flynn}}{{Brook} et~al.}{2004}]{Brook2004}
{Brook} C.~B.,  {Kawata} D.,  {Gibson} B.~K.,    {Flynn} C.,  2004, \mnras,
  349, 52

\bibitem[\protect\citeauthoryear{{Brough}, {Proctor}, {Forbes}, {Couch},
  {Collins}, {Burke} \& {Mann}}{{Brough} et~al.}{2007}]{Brough2007}
{Brough} S.,  {Proctor} R.,  {Forbes} D.~A.,  {Couch} W.~J.,  {Collins} C.~A.,
  {Burke} D.~J.,    {Mann} R.~G.,  2007, \mnras, 378, 1507

\bibitem[\protect\citeauthoryear{{Bruzual} \& {Charlot}}{{Bruzual} \&
  {Charlot}}{2003}]{Bruzual2003}
{Bruzual} G.,  {Charlot} S.,  2003, \mnras, 344, 1000

\bibitem[\protect\citeauthoryear{{Bryan}}{{Bryan}}{1999}]{Bryan99}
{Bryan} G.~L.,  1999, Comput.~Sci.~Eng., Vol.~1, No.~2, p.~46 - 53, 1, 46

\bibitem[\protect\citeauthoryear{{Bundy}, {Fukugita}, {Ellis}, {Kodama} \&
  {Conselice}}{{Bundy} et~al.}{2004}]{Bundy2004}
{Bundy} K.,  {Fukugita} M.,  {Ellis} R.~S.,  {Kodama} T.,    {Conselice} C.~J.,
   2004, \apjl, 601, L123

\bibitem[\protect\citeauthoryear{{Bundy}, {Fukugita}, {Ellis}, {Targett},
  {Belli} \& {Kodama}}{{Bundy} et~al.}{2009}]{Bundy2009}
{Bundy} K.,  {Fukugita} M.,  {Ellis} R.~S.,  {Targett} T.~A.,  {Belli} S.,
  {Kodama} T.,  2009, \apj, 697, 1369

\bibitem[\protect\citeauthoryear{{Carlberg}}{{Carlberg}}{1984}]{Carlberg1984}
{Carlberg} R.~G.,  1984, \apj, 286, 416

\bibitem[\protect\citeauthoryear{{Carrasco}, {Conselice} \&
  {Trujillo}}{{Carrasco} et~al.}{2010}]{Carrasco2010}
{Carrasco} E.~R.,  {Conselice} C.~J.,  {Trujillo} I.,  2010, \mnras, 405, 2253

\bibitem[\protect\citeauthoryear{{Cen}}{{Cen}}{2011a}]{Cen2011d}
{Cen} R.,  2011a, ArXiv e-prints

\bibitem[\protect\citeauthoryear{{Cen}}{{Cen}}{2011b}]{Cen2011}
{Cen} R.,  2011b, \apj, 741, 99

\bibitem[\protect\citeauthoryear{{Cen}}{{Cen}}{2011c}]{Cen2011c}
{Cen} R.,  2011c, \apjl, 742, L33

\bibitem[\protect\citeauthoryear{{Cen}}{{Cen}}{2011d}]{Cen2011b}
{Cen} R.,  2011d, ArXiv e-prints

\bibitem[\protect\citeauthoryear{{Cen}}{{Cen}}{2012}]{Cen2010}
{Cen} R.,  2012, \apj, 748, 121

\bibitem[\protect\citeauthoryear{{Cen}, {Kang}, {Ostriker} \& {Ryu}}{{Cen}
  et~al.}{1995}]{Cen95}
{Cen} R.,  {Kang} H.,  {Ostriker} J.~P.,    {Ryu} D.,  1995, \apj, 451, 436

\bibitem[\protect\citeauthoryear{{Cen}, {Nagamine} \& {Ostriker}}{{Cen}
  et~al.}{2005}]{Cen05}
{Cen} R.,  {Nagamine} K.,  {Ostriker} J.~P.,  2005, \apj, 635, 86

\bibitem[\protect\citeauthoryear{{Cen} \& {Ostriker}}{{Cen} \&
  {Ostriker}}{1992}]{Cen92}
{Cen} R.,  {Ostriker} J.~P.,  1992, \apjl, 399, L113

\bibitem[\protect\citeauthoryear{{Ceverino}, {Dekel} \& {Bournaud}}{{Ceverino}
  et~al.}{2010}]{Ceverino2010}
{Ceverino} D.,  {Dekel} A.,  {Bournaud} F.,  2010, \mnras, 404, 2151

\bibitem[\protect\citeauthoryear{{Coccato}, {Gerhard}, {Arnaboldi} \&
  {Ventimiglia}}{{Coccato} et~al.}{2011}]{Coccato2011}
{Coccato} L.,  {Gerhard} O.,  {Arnaboldi} M.,    {Ventimiglia} G.,  2011, \aap,
  533, A138

\bibitem[\protect\citeauthoryear{{Conselice}, {Yang} \& {Bluck}}{{Conselice}
  et~al.}{2009}]{Conselice2009}
{Conselice} C.~J.,  {Yang} C.,  {Bluck} A.~F.~L.,  2009, \mnras, 394, 1956

\bibitem[\protect\citeauthoryear{{Crain}, {Theuns}, {Dalla Vecchia}, {Eke},
  {Frenk}, {Jenkins}, {Kay}, {Peacock}, {Pearce}, {Schaye}, {Springel},
  {Thomas}, {White} \& {Wiersma}}{{Crain} et~al.}{2009}]{Crain2009}
{Crain} R.~A., et al. 2009, \mnras, 399, 1773

\bibitem[\protect\citeauthoryear{{Croton}, {Springel}, {White}, {De Lucia},
  {Frenk}, {Gao}, {Jenkins}, {Kauffmann}, {Navarro} \& {Yoshida}}{{Croton}
  et~al.}{2006}]{Croton06}
{Croton} D.~J., et al. 2006, \mnras, 365, 11

\bibitem[\protect\citeauthoryear{{Daddi}, {Renzini}, {Pirzkal}, {Cimatti},
  {Malhotra}, {Stiavelli}, {Xu}, {Pasquali}, {Rhoads}, {Brusa}, {di Serego
  Alighieri}, {Ferguson}, {Koekemoer}, {Moustakas}, {Panagia} \&
  {Windhorst}}{{Daddi} et~al.}{2005}]{Daddi2005}
{Daddi} E., et al. 2005, \apj, 626, 680

\bibitem[\protect\citeauthoryear{{Damjanov}, {McCarthy}, {Abraham},
  {Glazebrook}, {Yan}, {Mentuch}, {Le Borgne}, {Savaglio}, {Crampton},
  {Murowinski}, {Juneau}, {Carlberg}, {J{\o}rgensen}, {Roth}, {Chen} \&
  {Marzke}}{{Damjanov} et~al.}{2009}]{Damjanov09}
{Damjanov} I., et al. 2009, \apj, 695, 101

\bibitem[\protect\citeauthoryear{{Darg}, {Kaviraj}, {Lintott}, {Schawinski},
  {Sarzi}, {Bamford}, {Silk}, {Andreescu}, {Murray}, {Nichol}, {Raddick},
  {Slosar}, {Szalay}, {Thomas} \& {Vandenberg}}{{Darg} et~al.}{2010}]{Darg2010}
{Darg} D.~W., et al. 2010, \mnras, 401, 1552

\bibitem[\protect\citeauthoryear{{Davies}, {Sadler} \& {Peletier}}{{Davies}
  et~al.}{1993}]{Davies1993}
{Davies} R.~L.,  {Sadler} E.~M.,  {Peletier} R.~F.,  1993, \mnras, 262, 650

\bibitem[\protect\citeauthoryear{{de Ravel}, {Le F{\`e}vre}, {Tresse},
  {Bottini}, {Garilli}, {Le Brun}, {Maccagni}, {Scaramella}, {Scodeggio},
  {Vettolani}, {Zanichelli}, {Adami}, {Arnouts} et~al.,}{{de Ravel}
  et~al.}{2009}]{deRavel2009}
{de Ravel} L.,  {Le F{\`e}vre} O.,  {Tresse} L.,  {Bottini} D.,  {Garilli} B.,
  {Le Brun} V.,  {Maccagni} D.,  {Scaramella} R.,  {Scodeggio} M.,  {Vettolani}
  G.,  {Zanichelli} A.,  {Adami} C.,  {Arnouts} S.,    et~al., 2009, \aap, 498,
  379

\bibitem[\protect\citeauthoryear{{de Vaucouleurs}}{{de
  Vaucouleurs}}{1961}]{deVaucouleurs61}
{de Vaucouleurs} G.,  1961, \apjs, 5, 233

\bibitem[\protect\citeauthoryear{{Deason}, {Belokurov} \& {Evans}}{{Deason}
  et~al.}{2011}]{Deason2011}
{Deason} A.~J.,  {Belokurov} V.,  {Evans} N.~W.,  2011, \mnras, 411, 1480

\bibitem[\protect\citeauthoryear{{Dekel} \& {Birnboim}}{{Dekel} \&
  {Birnboim}}{2006}]{Dekel06}
{Dekel} A.,  {Birnboim} Y.,  2006, \mnras, 368, 2

\bibitem[\protect\citeauthoryear{{Dekel}, {Sari} \& {Ceverino}}{{Dekel}
  et~al.}{2009}]{Dekel09}
{Dekel} A.,  {Sari} R.,  {Ceverino} D.,  2009, \apj, 703, 785

\bibitem[\protect\citeauthoryear{{Eggen}, {Lynden-Bell} \& {Sandage}}{{Eggen}
  et~al.}{1962}]{ELS62}
{Eggen} O.~J.,  {Lynden-Bell} D.,  {Sandage} A.~R.,  1962, \apj, 136, 748

\bibitem[\protect\citeauthoryear{{Eisenhardt}, {De Propris}, {Gonzalez},
  {Stanford}, {Wang} \& {Dickinson}}{{Eisenhardt}
  et~al.}{2007}]{Eisenhardt2007}
{Eisenhardt} P.~R.,  {De Propris} R.,  {Gonzalez} A.~H.,  {Stanford} S.~A.,
  {Wang} M.,    {Dickinson} M.,  2007, \apjs, 169, 225

\bibitem[\protect\citeauthoryear{{Eisenstein} \& {Hu}}{{Eisenstein} \&
  {Hu}}{1999}]{Eisenstein99}
{Eisenstein} D.~J.,  {Hu} W.,  1999, \apj, 511, 5

\bibitem[\protect\citeauthoryear{{Elmegreen}, {Elmegreen}, {Fernandez} \&
  {Lemonias}}{{Elmegreen} et~al.}{2009}]{Elmegreen09}
{Elmegreen} B.~G.,  {Elmegreen} D.~M.,  {Fernandez} M.~X.,    {Lemonias} J.~J.,
   2009, \apj, 692, 12

\bibitem[\protect\citeauthoryear{{Evans} II, {Dunham}, {J{\o}rgensen}, {Enoch},
  {Mer{\'{\i}}n}, {van Dishoeck}, {Alcal{\'a}}, {Myers}, {Stapelfeldt},
  {Huard}, {Allen} et~al.,}{{Evans} et~al.}{2009}]{Evans09}
{Evans} II N.~J., et al. 2009,   \apjs, 181, 321

\bibitem[\protect\citeauthoryear{{Faber}, {Burstein} \& {Dressler}}{{Faber}
  et~al.}{1977}]{Faber1977}
{Faber} S.~M.,  {Burstein} D.,  {Dressler} A.,  1977, \aj, 82, 941

\bibitem[\protect\citeauthoryear{{Fakhouri} \& {Ma}}{{Fakhouri} \&
  {Ma}}{2008}]{Fakhouri2008}
{Fakhouri} O.,  {Ma} C.-P.,  2008, \mnras, 386, 577

\bibitem[\protect\citeauthoryear{{Fakhouri}, {Ma} \&
  {Boylan-Kolchin}}{{Fakhouri} et~al.}{2010}]{Fakhouri2010}
{Fakhouri} O.,  {Ma} C.-P.,  {Boylan-Kolchin} M.,  2010, \mnras, 406, 2267

\bibitem[\protect\citeauthoryear{{Font}, {Johnston}, {Bullock} \&
  {Robertson}}{{Font} et~al.}{2006a}]{Font06b}
{Font} A.~S.,  {Johnston} K.~V.,  {Bullock} J.~S.,    {Robertson} B.~E.,
  2006a, \apj, 638, 585

\bibitem[\protect\citeauthoryear{{Font}, {Johnston}, {Bullock} \&
  {Robertson}}{{Font} et~al.}{2006b}]{Font06a}
{Font} A.~S.,  {Johnston} K.~V.,  {Bullock} J.~S.,    {Robertson} B.~E.,
  2006b, \apj, 646, 886

\bibitem[\protect\citeauthoryear{{Font}, {McCarthy}, {Crain}, {Theuns},
  {Schaye}, {Wiersma} \& {Dalla Vecchia}}{{Font} et~al.}{2011}]{Font2011}
{Font} A.~S.,  {McCarthy} I.~G.,  {Crain} R.~A.,  {Theuns} T.,  {Schaye} J.,
  {Wiersma} R.~P.~C.,    {Dalla Vecchia} C.,  2011, \mnras, 416, 2802

\bibitem[\protect\citeauthoryear{{Forbes}, {Spitler}, {Strader}, {Romanowsky},
  {Brodie} \& {Foster}}{{Forbes} et~al.}{2011}]{Forbes2011}
{Forbes} D.~A.,  {Spitler} L.~R.,  {Strader} J.,  {Romanowsky} A.~J.,  {Brodie}
  J.~P.,    {Foster} C.,  2011, \mnras, 413, 2943

\bibitem[\protect\citeauthoryear{{Foster}, {Proctor}, {Forbes}, {Spolaor},
  {Hopkins} \& {Brodie}}{{Foster} et~al.}{2009}]{Foster2009}
{Foster} C.,  {Proctor} R.~N.,  {Forbes} D.~A.,  {Spolaor} M.,  {Hopkins}
  P.~F.,    {Brodie} J.~P.,  2009, \mnras, 400, 2135

\bibitem[\protect\citeauthoryear{{Fukugita}, {Ichikawa}, {Gunn}, {Doi},
  {Shimasaku} \& {Schneider}}{{Fukugita} et~al.}{1996}]{Fukugita1996}
{Fukugita} M.,  {Ichikawa} T.,  {Gunn} J.~E.,  {Doi} M.,  {Shimasaku} K.,
  {Schneider} D.~P.,  1996, \aj, 111, 1748

\bibitem[\protect\citeauthoryear{{Fukugita} \& {Peebles}}{{Fukugita} \&
  {Peebles}}{2004}]{Fukugita2004}
{Fukugita} M.,  {Peebles} P.~J.~E.,  2004, \apj, 616, 643

\bibitem[\protect\citeauthoryear{{Gallagher} III \& {Ostriker}}{{Gallagher} \&
  {Ostriker}}{1972}]{Gallagher72}
{Gallagher} III J.~S.,  {Ostriker} J.~P.,  1972, \aj, 77, 288

\bibitem[\protect\citeauthoryear{{Gallazzi}, {Brinchmann}, {Charlot} \&
  {White}}{{Gallazzi} et~al.}{2008}]{Gallazzi2008}
{Gallazzi} A.,  {Brinchmann} J.,  {Charlot} S.,    {White} S.~D.~M.,  2008,
  \mnras, 383, 1439

\bibitem[\protect\citeauthoryear{{Gallazzi}, {Charlot}, {Brinchmann}, {White}
  \& {Tremonti}}{{Gallazzi} et~al.}{2005}]{Gallazzi2005}
{Gallazzi} A.,  {Charlot} S.,  {Brinchmann} J.,  {White} S.~D.~M.,
  {Tremonti} C.~A.,  2005, \mnras, 362, 41

\bibitem[\protect\citeauthoryear{{Gonzalez-Perez}, {Castander} \&
  {Kauffmann}}{{Gonzalez-Perez} et~al.}{2011}]{GonzalezPerez2011}
{Gonzalez-Perez} V.,  {Castander} F.~J.,  {Kauffmann} G.,  2011, \mnras, 411,
  1151

\bibitem[\protect\citeauthoryear{{Gottl{\"o}ber}, {Klypin} \&
  {Kravtsov}}{{Gottl{\"o}ber} et~al.}{2001}]{Gottlober2001}
{Gottl{\"o}ber} S.,  {Klypin} A.,  {Kravtsov} A.~V.,  2001, \apj, 546, 223

\bibitem[\protect\citeauthoryear{{Greene}, {Murphy}, {Comerford}, {Gebhardt} \&
  {Adams}}{{Greene} et~al.}{2012}]{Greene2012}
{Greene} J.~E.,  {Murphy} J.~D.,  {Comerford} J.~M.,  {Gebhardt} K.,    {Adams}
  J.~J.,  2012, ArXiv e-prints

\bibitem[\protect\citeauthoryear{{Guo}, {White}, {Li} \&
  {Boylan-Kolchin}}{{Guo} et~al.}{2010}]{Guo2010}
{Guo} Q.,  {White} S.,  {Li} C.,    {Boylan-Kolchin} M.,  2010, \mnras, 404,
  1111

\bibitem[\protect\citeauthoryear{{Guo}, {Giavalisco}, {Cassata}, {Ferguson},
  {Dickinson}, {Renzini}, {Koekemoer}, {Grogin}, {Papovich}, {Tundo},
  {Fontana}, {Lotz} \& {Salimbeni}}{{Guo} et~al.}{2011}]{Guo2011}
{Guo} Y., et al. 2011, \apj, 735, 18

\bibitem[\protect\citeauthoryear{{Haardt} \& {Madau}}{{Haardt} \&
  {Madau}}{1996}]{Haardt96}
{Haardt} F.,  {Madau} P.,  1996, \apj, 461, 20

\bibitem[\protect\citeauthoryear{{Hopkins}, {Bundy}, {Croton}, {Hernquist},
  {Keres}, {Khochfar}, {Stewart}, {Wetzel} \& {Younger}}{{Hopkins}
  et~al.}{2010}]{Hopkins10}
{Hopkins} P.~F., et al. 2010, \apj,   715, 202

\bibitem[\protect\citeauthoryear{{Hopkins}, {Bundy}, {Murray}, {Quataert},
  {Lauer} \& {Ma}}{{Hopkins} et~al.}{2009}]{Hopkins09a}
{Hopkins} P.~F.,  {Bundy} K.,  {Murray} N.,  {Quataert} E.,  {Lauer} T.~R.,
  {Ma} C.-P.,  2009, \mnras, 398, 898

\bibitem[\protect\citeauthoryear{{Jogee}, {Miller}, {Penner}, {Skelton},
  {Conselice}, {Somerville}, {Bell}, {Zheng}, {Rix}, {Robaina}, {Barazza}
  et~al.,}{{Jogee} et~al.}{2009}]{Jogee2009}
{Jogee} S., et al. 2009, \apj, 697, 1971

\bibitem[\protect\citeauthoryear{{Johansson}, {Naab} \& {Ostriker}}{{Johansson}
  et~al.}{2012}]{Johansson2012}
{Johansson} P.~H.,  {Naab} T.,  {Ostriker} J.~P.,  2012, ArXiv e-prints

\bibitem[\protect\citeauthoryear{{Johnston}, {Bullock}, {Sharma}, {Font},
  {Robertson} \& {Leitner}}{{Johnston} et~al.}{2008}]{Johnston08}
{Johnston} K.~V.,  {Bullock} J.~S.,  {Sharma} S.,  {Font} A.,  {Robertson}
  B.~E.,    {Leitner} S.~N.,  2008, \apj, 689, 936

\bibitem[\protect\citeauthoryear{{Joung}, {Cen} \& {Bryan}}{{Joung}
  et~al.}{2009}]{Joung09}
{Joung} M.~R.,  {Cen} R.,  {Bryan} G.~L.,  2009, \apjl, 692, L1

\bibitem[\protect\citeauthoryear{{Kartaltepe}, {Sanders}, {Le Floc'h},
  {Frayer}, {Aussel}, {Arnouts}, {Ilbert}, {Salvato}, {Scoville}, {Surace},
  {Yan}, {Capak}, {Caputi} et~al.,}{{Kartaltepe} et~al.}{2010}]{Kartaltepe2010}
{Kartaltepe} J.~S., et al. 2010, \apj,   721, 98

\bibitem[\protect\citeauthoryear{{Kartaltepe}, {Sanders}, {Scoville},
  {Calzetti}, {Capak}, {Koekemoer}, {Mobasher}, {Murayama}, {Salvato}, {Sasaki}
  \& {Taniguchi}}{{Kartaltepe} et~al.}{2007}]{Kartaltepe2007}
{Kartaltepe} J.~S., et al. 2007, \apjs, 172, 320

\bibitem[\protect\citeauthoryear{{Kauffmann}, {White} \&
  {Guiderdoni}}{{Kauffmann} et~al.}{1993}]{Kauffmann1993}
{Kauffmann} G.,  {White} S.~D.~M.,  {Guiderdoni} B.,  1993, \mnras, 264, 201

\bibitem[\protect\citeauthoryear{{Kennicutt}
  Jr.}{{Kennicutt}}{1998}]{Kennicutt98}
{Kennicutt} Jr. R.~C.,  1998, \apj, 498, 541

\bibitem[\protect\citeauthoryear{{Komatsu}, {Smith}, {Dunkley}, {Bennett},
  {Gold}, {Hinshaw}, {Jarosik}, {Larson}, {Nolta}, {Page}, {Spergel}, {Halpern}
  et~al.,}{{Komatsu} et~al.}{2011}]{Komatsu11}
{Komatsu} E., et al. 2011, \apjs, 192, 18

\bibitem[\protect\citeauthoryear{{Krumholz} \& {McKee}}{{Krumholz} \&
  {McKee}}{2005}]{Krumholz05}
{Krumholz} M.~R.,  {McKee} C.~F.,  2005, \apj, 630, 250

\bibitem[\protect\citeauthoryear{{Kuntschner}, {Emsellem}, {Bacon},
  {Cappellari}, {Davies}, {de Zeeuw}, {Falc{\'o}n-Barroso}, {Krajnovi{\'c}},
  {McDermid}, {Peletier}, {Sarzi}, {Shapiro}, {van den Bosch} \& {van de
  Ven}}{{Kuntschner} et~al.}{2010}]{Kuntschner2010}
{Kuntschner} H., et al. 2010, \mnras, 408, 97

\bibitem[\protect\citeauthoryear{{La Barbera}, {Ferreras}, {de Carvalho},
  {Lopes}, {Pasquali}, {de la Rosa} \& {De Lucia}}{{La Barbera}
  et~al.}{2011}]{LaBarbera2011}
{La Barbera} F.,  {Ferreras} I.,  {de Carvalho} R.~R.,  {Lopes} P.~A.~A.,
  {Pasquali} A.,  {de la Rosa} I.~G.,    {De Lucia} G.,  2011, \apjl, 740, L41

\bibitem[\protect\citeauthoryear{{Larson}}{{Larson}}{1975}]{Larson1975}
{Larson} R.~B.,  1975, \mnras, 173, 671

\bibitem[\protect\citeauthoryear{{Le F{\`e}vre}, {Abraham}, {Lilly}, {Ellis},
  {Brinchmann}, {Schade}, {Tresse}, {Colless}, {Crampton}, {Glazebrook},
  {Hammer} \& {Broadhurst}}{{Le F{\`e}vre} et~al.}{2000}]{LeFevre00}
{Le F{\`e}vre} O., et al. 2000, \mnras, 311, 565

\bibitem[\protect\citeauthoryear{{Leauthaud}, {George}, {Behroozi}, {Bundy},
  {Tinker}, {Wechsler}, {Conroy}, {Finoguenov} \& {Tanaka}}{{Leauthaud}
  et~al.}{2012}]{Leauthaud2012}
{Leauthaud} A., et al. 2012,   \apj, 746, 95

\bibitem[\protect\citeauthoryear{{Li} \& {White}}{{Li} \&
  {White}}{2009}]{Li2009}
{Li} C.,  {White} S.~D.~M.,  2009, \mnras, 398, 2177

\bibitem[\protect\citeauthoryear{{Lin}, {Cooper}, {Jian}, {Koo}, {Patton},
  {Yan}, {Willmer}, {Coil}, {Chiueh}, {Croton}, {Gerke}, {Lotz}, {Guhathakurta}
  \& {Newman}}{{Lin} et~al.}{2010}]{Lin2010}
{Lin} L., et al. 2010,   \apj, 718, 1158

\bibitem[\protect\citeauthoryear{{Lin}, {Koo}, {Willmer}, {Patton},
  {Conselice}, {Yan}, {Coil}, {Cooper}, {Davis}, {Faber}, {Gerke},
  {Guhathakurta} \& {Newman}}{{Lin} et~al.}{2004}]{Lin2004}
{Lin} L., et al. 2004, \apjl,   617, L9

\bibitem[\protect\citeauthoryear{{L{\'o}pez-Sanjuan}, {Balcells},
  {P{\'e}rez-Gonz{\'a}lez}, {Barro}, {Garc{\'{\i}}a-Dab{\'o}}, {Gallego} \&
  {Zamorano}}{{L{\'o}pez-Sanjuan} et~al.}{2009}]{LopezSanjuan2009}
{L{\'o}pez-Sanjuan} C.,  {Balcells} M.,  {P{\'e}rez-Gonz{\'a}lez} P.~G.,
  {Barro} G.,  {Garc{\'{\i}}a-Dab{\'o}} C.~E.,  {Gallego} J.,    {Zamorano} J.,
   2009, \aap, 501, 505

\bibitem[\protect\citeauthoryear{{Lotz}, {Davis}, {Faber}, {Guhathakurta},
  {Gwyn}, {Huang}, {Koo}, {Le Floc'h}, {Lin}, {Newman}, {Noeske}, {Papovich}
  et~al.,}{{Lotz} et~al.}{2008}]{Lotz2008}
{Lotz} J.~M., et al. 2008, \apj, 672, 177

\bibitem[\protect\citeauthoryear{{Lotz}, {Jonsson}, {Cox}, {Croton}, {Primack},
  {Somerville} \& {Stewart}}{{Lotz} et~al.}{2011}]{Lotz2011}
{Lotz} J.~M.,  {Jonsson} P.,  {Cox} T.~J.,  {Croton} D.,  {Primack} J.~R.,
  {Somerville} R.~S.,    {Stewart} K.,  2011, \apj, 742, 103

\bibitem[\protect\citeauthoryear{{Masjedi}, {Hogg} \& {Blanton}}{{Masjedi}
  et~al.}{2008}]{Masjedi08}
{Masjedi} M.,  {Hogg} D.~W.,  {Blanton} M.~R.,  2008, \apj, 679, 260

\bibitem[\protect\citeauthoryear{{McIntosh}, {Guo}, {Hertzberg}, {Katz}, {Mo},
  {van den Bosch} \& {Yang}}{{McIntosh} et~al.}{2008}]{McIntosh2008}
{McIntosh} D.~H.,  {Guo} Y.,  {Hertzberg} J.,  {Katz} N.,  {Mo} H.~J.,  {van
  den Bosch} F.~C.,    {Yang} X.,  2008, \mnras, 388, 1537

\bibitem[\protect\citeauthoryear{{Mehlert}, {Thomas}, {Saglia}, {Bender} \&
  {Wegner}}{{Mehlert} et~al.}{2003}]{Mehlert2003}
{Mehlert} D.,  {Thomas} D.,  {Saglia} R.~P.,  {Bender} R.,    {Wegner} G.,
  2003, \aap, 407, 423

\bibitem[\protect\citeauthoryear{{Moore}, {Katz} \& {Lake}}{{Moore}
  et~al.}{1996}]{Moore96}
{Moore} B.,  {Katz} N.,  {Lake} G.,  1996, \apj, 457, 455

\bibitem[\protect\citeauthoryear{{Moster}, {Somerville}, {Maulbetsch}, {van den
  Bosch}, {Macci{\`o}}, {Naab} \& {Oser}}{{Moster} et~al.}{2010}]{Moster2010}
{Moster} B.~P.,  {Somerville} R.~S.,  {Maulbetsch} C.,  {van den Bosch} F.~C.,
  {Macci{\`o}} A.~V.,  {Naab} T.,    {Oser} L.,  2010, \apj, 710, 903

\bibitem[\protect\citeauthoryear{{Naab}, {Johansson} \& {Ostriker}}{{Naab}
  et~al.}{2009}]{Naab09}
{Naab} T.,  {Johansson} P.~H.,  {Ostriker} J.~P.,  2009, \apjl, 699, L178

\bibitem[\protect\citeauthoryear{{Naab}, {Johansson}, {Ostriker} \&
  {Efstathiou}}{{Naab} et~al.}{2007}]{Naab07}
{Naab} T.,  {Johansson} P.~H.,  {Ostriker} J.~P.,    {Efstathiou} G.,  2007,
  \apj, 658, 710

\bibitem[\protect\citeauthoryear{{Newman}, {Ellis}, {Bundy} \& {Treu}}{{Newman}
  et~al.}{2012}]{Newman2012}
{Newman} A.~B.,  {Ellis} R.~S.,  {Bundy} K.,    {Treu} T.,  2012, \apj, 746,
  162

\bibitem[\protect\citeauthoryear{{Nissen} \& {Schuster}}{{Nissen} \&
  {Schuster}}{2010}]{Nissen2010}
{Nissen} P.~E.,  {Schuster} W.~J.,  2010, \aap, 511, L10

\bibitem[\protect\citeauthoryear{{Oser}, {Naab}, {Ostriker} \&
  {Johansson}}{{Oser} et~al.}{2012}]{Oser11}
{Oser} L.,  {Naab} T.,  {Ostriker} J.~P.,    {Johansson} P.~H.,  2012, \apj,
  744, 63

\bibitem[\protect\citeauthoryear{{Oser}, {Ostriker}, {Naab}, {Johansson} \&
  {Burkert}}{{Oser} et~al.}{2010}]{Oser10}
{Oser} L.,  {Ostriker} J.~P.,  {Naab} T.,  {Johansson} P.~H.,    {Burkert} A.,
  2010, \apj, 725, 2312

\bibitem[\protect\citeauthoryear{{O'Shea}, {Bryan}, {Bordner}, {Norman},
  {Abel}, {Harkness} \& {Kritsuk}}{{O'Shea} et~al.}{2004}]{OShea04}
{O'Shea} B.~W.,  {Bryan} G.,  {Bordner} J.,  {Norman} M.~L.,  {Abel} T.,
  {Harkness} R.,    {Kritsuk} A.,  2004, ArXiv Astrophysics e-prints

\bibitem[\protect\citeauthoryear{{Ostriker} \& {Hausman}}{{Ostriker} \&
  {Hausman}}{1977}]{OstrikerHausman77}
{Ostriker} J.~P.,  {Hausman} M.~A.,  1977, \apjl, 217, L125

\bibitem[\protect\citeauthoryear{{Peebles}}{{Peebles}}{2001}]{Peebles2001}
{Peebles} P.~J.~E.,  2001, \apj, 557, 495

\bibitem[\protect\citeauthoryear{{Rawle}, {Smith}, {Lucey} \&
  {Swinbank}}{{Rawle} et~al.}{2008}]{Rawle2008}
{Rawle} T.~D.,  {Smith} R.~J.,  {Lucey} J.~R.,    {Swinbank} A.~M.,  2008,
  \mnras, 389, 1891

\bibitem[\protect\citeauthoryear{{Raymond}, {Cox} \& {Smith}}{{Raymond}
  et~al.}{1976}]{Raymond1976}
{Raymond} J.~C.,  {Cox} D.~P.,  {Smith} B.~W.,  1976, \apj, 204, 290

\bibitem[\protect\citeauthoryear{{Robertson}, {Bullock}, {Font}, {Johnston} \&
  {Hernquist}}{{Robertson} et~al.}{2005}]{Robertson05}
{Robertson} B.,  {Bullock} J.~S.,  {Font} A.~S.,  {Johnston} K.~V.,
  {Hernquist} L.,  2005, \apj, 632, 872

\bibitem[\protect\citeauthoryear{{Scannapieco}, {Wadepuhl}, {Parry}, {Navarro},
  {Jenkins}, {Springel}, {Teyssier}, {Carlson}, {Couchman}, {Crain}, {Dalla
  Vecchia}, {Frenk}, {Kobayashi} et~al.,}{{Scannapieco}
  et~al.}{2011}]{Scannapieco2011}
{Scannapieco} C., et al. 2011, ArXiv e-prints

\bibitem[\protect\citeauthoryear{{Searle} \& {Zinn}}{{Searle} \&
  {Zinn}}{1978}]{SearleZinn78}
{Searle} L.,  {Zinn} R.,  1978, \apj, 225, 357

\bibitem[\protect\citeauthoryear{{Spinrad} \& {Taylor}}{{Spinrad} \&
  {Taylor}}{1971}]{Spinrad1971}
{Spinrad} H.,  {Taylor} B.~J.,  1971, \apjs, 22, 445

\bibitem[\protect\citeauthoryear{{Spolaor} et~al.,}{{Spolaor}
  et~al.}{2010}]{Spolaor2010}
{Spolaor} M.,  et~al., 2010, \mnras, 408, 272

\bibitem[\protect\citeauthoryear{{Springel}, {White}, {Jenkins}, {Frenk},
  {Yoshida}, {Gao}, {Navarro}, {Thacker}, {Croton}, {Helly}, {Peacock}, {Cole},
  {Thomas}, {Couchman}, {Evrard}, {Colberg} \& {Pearce}}{{Springel}
  et~al.}{2005}]{Springel2005}
{Springel} V., et al. 2005, \nat, 435, 629

\bibitem[\protect\citeauthoryear{{Suh}, {Jeong}, {Oh}, {Yi}, {Ferreras} \&
  {Schawinski}}{{Suh} et~al.}{2010}]{Suh2010}
{Suh} H.,  {Jeong} H.,  {Oh} K.,  {Yi} S.~K.,  {Ferreras} I.,    {Schawinski}
  K.,  2010, \apjs, 187, 374

\bibitem[\protect\citeauthoryear{{Tal} \& {van Dokkum}}{{Tal} \& {van
  Dokkum}}{2011}]{Tal2011}
{Tal} T.,  {van Dokkum} P.~G.,  2011, \apj, 731, 89

\bibitem[\protect\citeauthoryear{{Tal}, {Wake}, {van Dokkum}, {van den Bosch},
  {Schneider}, {Brinkmann} \& {Weaver}}{{Tal} et~al.}{2012}]{Tal2012}
{Tal} T.,  {Wake} D.~A.,  {van Dokkum} P.~G.,  {van den Bosch} F.~C.,
  {Schneider} D.~P.,  {Brinkmann} J.,    {Weaver} B.~A.,  2012, \apj, 746, 138

\bibitem[\protect\citeauthoryear{{Thomas}, {Maraston}, {Bender} \& {Mendes de
  Oliveira}}{{Thomas} et~al.}{2005}]{Thomas05}
{Thomas} D.,  {Maraston} C.,  {Bender} R.,    {Mendes de Oliveira} C.,  2005,
  \apj, 621, 673

\bibitem[\protect\citeauthoryear{{Tissera}, {White} \& {Scannapieco}}{{Tissera}
  et~al.}{2012}]{Tissera2012}
{Tissera} P.~B.,  {White} S.~D.~M.,  {Scannapieco} C.,  2012, \mnras, 420, 255

\bibitem[\protect\citeauthoryear{{Tonnesen} \& {Cen}}{{Tonnesen} \&
  {Cen}}{2011}]{Tonnesen2011}
{Tonnesen} S.,  {Cen} R.,  2011, ArXiv e-prints

\bibitem[\protect\citeauthoryear{{Tortora}, {Napolitano}, {Cardone},
  {Capaccioli}, {Jetzer} \& {Molinaro}}{{Tortora} et~al.}{2010}]{Tortora2010}
{Tortora} C.,  {Napolitano} N.~R.,  {Cardone} V.~F.,  {Capaccioli} M.,
  {Jetzer} P.,    {Molinaro} R.,  2010, \mnras, 407, 144

\bibitem[\protect\citeauthoryear{{Trujillo}, {Conselice}, {Bundy}, {Cooper},
  {Eisenhardt} \& {Ellis}}{{Trujillo} et~al.}{2007}]{Trujillo07}
{Trujillo} I.,  {Conselice} C.~J.,  {Bundy} K.,  {Cooper} M.~C.,  {Eisenhardt}
  P.,    {Ellis} R.~S.,  2007, \mnras, 382, 109

\bibitem[\protect\citeauthoryear{{Unavane}, {Wyse} \& {Gilmore}}{{Unavane}
  et~al.}{1996}]{Unavane96}
{Unavane} M.,  {Wyse} R.~F.~G.,  {Gilmore} G.,  1996, \mnras, 278, 727

\bibitem[\protect\citeauthoryear{{van de Sande}, {Kriek}, {Franx}, {van
  Dokkum}, {Bezanson}, {Whitaker}, {Brammer}, {Labb{\'e}}, {Groot} \&
  {Kaper}}{{van de Sande} et~al.}{2011}]{vandeSande11}
{van de Sande} J.,  {Kriek} M.,  {Franx} M.,  {van Dokkum} P.~G.,  {Bezanson}
  R.,  {Whitaker} K.~E.,  {Brammer} G.,  {Labb{\'e}} I.,  {Groot} P.~J.,
  {Kaper} L.,  2011, \apjl, 736, L9

\bibitem[\protect\citeauthoryear{{van der Wel}, {Holden}, {Zirm}, {Franx},
  {Rettura}, {Illingworth} \& {Ford}}{{van der Wel}
  et~al.}{2008}]{vanderWel2008}
{van der Wel} A.,  {Holden} B.~P.,  {Zirm} A.~W.,  {Franx} M.,  {Rettura} A.,
  {Illingworth} G.~D.,    {Ford} H.~C.,  2008, \apj, 688, 48

\bibitem[\protect\citeauthoryear{{van der Wel}, {Rix}, {Wuyts}, {McGrath},
  {Koekemoer}, {Bell}, {Holden}, {Robaina} \& {McIntosh}}{{van der Wel}
  et~al.}{2011}]{vanderWel2011}
{van der Wel} A.,  {Rix} H.-W.,  {Wuyts} S.,  {McGrath} E.~J.,  {Koekemoer}
  A.~M.,  {Bell} E.~F.,  {Holden} B.~P.,  {Robaina} A.~R.,    {McIntosh} D.~H.,
   2011, \apj, 730, 38

\bibitem[\protect\citeauthoryear{{van Dokkum} \& {Brammer}}{{van Dokkum} \&
  {Brammer}}{2010}]{vanDokkum2010}
{van Dokkum} P.~G.,  {Brammer} G.,  2010, \apjl, 718, L73

\bibitem[\protect\citeauthoryear{{van Dokkum}, {Franx}, {Kriek}, {Holden},
  {Illingworth}, {Magee}, {Bouwens}, {Marchesini}, {Quadri}, {Rudnick},
  {Taylor} \& {Toft}}{{van Dokkum} et~al.}{2008}]{vanDokkum08}
{van Dokkum} P.~G.,  {Franx} M.,  {Kriek} M.,  {Holden} B.,  {Illingworth}
  G.~D.,  {Magee} D.,  {Bouwens} R.,  {Marchesini} D.,  {Quadri} R.,  {Rudnick}
  G.,  {Taylor} E.~N.,    {Toft} S.,  2008, \apjl, 677, L5

\bibitem[\protect\citeauthoryear{{White}}{{White}}{1976}]{White76}
{White} S.~D.~M.,  1976, \mnras, 177, 717

\bibitem[\protect\citeauthoryear{{White} \& {Rees}}{{White} \&
  {Rees}}{1978}]{White1978}
{White} S.~D.~M.,  {Rees} M.~J.,  1978, \mnras, 183, 341

\bibitem[\protect\citeauthoryear{{Wiersma}, {Schaye} \& {Theuns}}{{Wiersma}
  et~al.}{2011}]{Wiersma2011}
{Wiersma} R.~P.~C.,  {Schaye} J.,  {Theuns} T.,  2011, \mnras, 415, 353

\bibitem[\protect\citeauthoryear{{Woodward} \& {Colella}}{{Woodward} \&
  {Colella}}{1984}]{Woodward1984}
{Woodward} P.,  {Colella} P.,  1984, Journal of Computational Physics, 54, 115

\bibitem[\protect\citeauthoryear{{Zolotov}, {Willman}, {Brooks}, {Governato},
  {Brook}, {Hogg}, {Quinn} \& {Stinson}}{{Zolotov} et~al.}{2009}]{Zolotov2009}
{Zolotov} A., et al. 2009, \apj, 702, 1058

\bibitem[\protect\citeauthoryear{{Zolotov}, {Willman}, {Brooks}, {Governato},
  {Hogg}, {Shen} \& {Wadsley}}{{Zolotov} et~al.}{2010}]{Zolotov2010}
{Zolotov} A.,  {Willman} B.,  {Brooks} A.~M.,  {Governato} F.,  {Hogg} D.~W.,
  {Shen} S.,    {Wadsley} J.,  2010, \apj, 721, 738

\end{thebibliography}

\label{lastpage}

\end{document}